\documentclass[aap,MSNbibl,seceqn,dvips]{arximspdf}
\usepackage[mathscr]{eucal}
\usepackage{mathrsfs}
\usepackage{graphicx}

\doi{10.1214/12-AAP845} 
\volume{23}
\issue{1}
\pubyear{2013}
\firstpage{348}
\lastpage{385}

\makeatletter

\newcommand{\mmbox}[1]{\mbox{\fontsize{8.36pt}{11pt}\selectfont{#1}}}

\newtheorem{theorem}{Theorem}[section]
\newtheorem{proposition}[theorem]{Proposition}
\newtheorem{lemma}[theorem]{Lemma}

\newproclaim{assumption}[theorem]{Assumption}
\newproclaim{remark}[theorem]{Remark}

\newcommand{\Def}{\stackrel{\mathrm{def}}{=}}
\newcommand{\Deff}{\stackrel{\mathit{def}}{=}}

\newcommand{\R}{\mathbb{R}}
\newcommand{\N}{\mathbb{N}}

\newcommand{\KK}{K}

\newcommand{\vrho}{\varrho}
\newcommand{\eps}{\varepsilon}

\newcommand{\Borel}{\mathscr{B}}
\newcommand{\Pspace}{\mathscr{P}}
\newcommand{\BP}{\mathbb{P}}
\newcommand{\BE}{\mathbb{E}}
\newcommand{\filt}{\mathscr{F}}
\newcommand{\gilt}{\mathscr{G}}

\newcommand{\ee}{\mathfrak{e}}
\newcommand{\genA}{\mathcal{A}}
\newcommand{\genL}{\mathcal{L}}
\newcommand{\jump}{\mathcal{J}}

\newcommand{\pt}{\star}
\newcommand{\SSS}{\mathcal{S}}

\newcommand{\PP}{\mathcal{P}}
\newcommand{\compact}{\mathcal{K}}
\newcommand{\NN}{{N,n}}
\newcommand{\mart}{\mathcal{M}}
\newcommand{\dfi}{\mbox{\textsc{m}}}
\newcommand{\QQ}{\mathcal{Q}}

\makeatother

\begin{document}
\begin{frontmatter}

\title{Default clustering in large portfolios: Typical~events}
\runtitle{Default clustering in large portfolios: Typical events}

\begin{aug}
\author[A]{\fnms{Kay} \snm{Giesecke}\ead[label=e1]{giesecke@stanford.edu}},
\author[B]{\fnms{Konstantinos} \snm{Spiliopoulos}\corref{}\ead[label=e2]{kspiliop@dam.brown.edu}}\\
\and
\author[C]{\fnms{Richard B.} \snm{Sowers}\ead[label=e3]{r-sowers@illinois.edu}}
\runauthor{K. Giesecke, K. Spiliopoulos and R. B. Sowers}
\affiliation{Stanford University, Brown University and
University of Illinois at~Urbana,~Champaign}
\address[A]{K. Giesecke \\
Department of Management Science\\
\quad and Engineering\\
Stanford University \\
Stanford, California 94305-4026 \\
USA \\
\printead{e1}}
\address[B]{K. Spiliopoulos \\
Division of Applied Mathematics \\
Brown University \\
Providence, Rhode Island 029126\\
USA \\
\printead{e2}}
\address[C]{R. B. Sowers \\
Department of Mathematics\\
University of Illinois\\
\quad at Urbana, Champaign\\
Urbana, Illinois 61801 \\
USA \\
\printead{e3}} 
\end{aug}

\received{\smonth{1} \syear{2011}}
\revised{\smonth{1} \syear{2012}}

%
\begin{abstract}
We develop a dynamic point process model of correlated default timing
in a portfolio of firms, and analyze typical default profiles in the
limit as the size of the pool grows. In our model, a firm defaults at
a stochastic intensity that is influenced by an idiosyncratic risk
process, a systematic risk process common to all firms, and past
defaults. We prove a law of large numbers for the default rate in the
pool, which describes the ``typical'' behavior of defaults.
\end{abstract}

%

\end{frontmatter}

\section{Introduction}\label{SIntro}
The financial crisis of 2007--09 has made clear the need to better
understand the diversification of risk in financial systems with
interacting entities. Prior to the crisis, the common belief was that
risk had been diversified away by using the tools of structured
finance. As it turned out, the correlation between assets was much
larger than supposed. The collapse fed on itself and created a spiral.

We study the behavior of defaults in a large portfolio of interacting
firms. We develop a dynamic point process model of correlated default
timing, and then analyze typical default profiles in the limit as the
number of constituent firms grows. Our empirically motivated model
incorporates two distinct sources of default clustering. First, the
firms are exposed to a risk factor process that is common to all
entities in the pool. Variations in this systematic risk factor
generate correlated movements in firms' conditional default
probabilities. Das, Duffie, Kapadia and Saita~\cite{ddk} show
that this mechanism is responsible for a large amount of corporate
default clustering in the U.S. Second, a default has a contagious
impact on the health of other firms. This impact fades away with time.
Azizpour, Giesecke and Schwenkler \cite
{azizpour-giesecke-schwenkler} provide statistical evidence for the
presence of such self-exciting effects in U.S. corporate defaults,
after controlling for the exposure of firms to systematic risk factors.

More precisely, we assume that a firm defaults at an intensity, or
conditional arrival rate, which follows a mean-reverting jump-diffusion
process that is driven by several terms. The first term, a square root
diffusion, represents an independent, firm-specific source of risk. The
second term is a systematic risk factor that influences all firms, and
that generates diffusive correlation between the intensities. For
simplicity, we take this systematic risk factor to be an
Ornstein--Uhlenbeck process. The third term affecting the intensity is
the default rate in the pool. Defaults cause jumps in the intensity;
they are common to all surviving firms. We thus have two sources of
correlation between the firms: the dependence on the systematic risk
factor and the influence of past defaults. While this formulation
parsimoniously captures several of the sources of default correlation
identified in empirical research, the intricate event dependence
structure presents a challenge for the mathematical analysis of the system.

Our goal is to understand the behavior of the default rate in the
portfolio in the limit as the number of firms in the pool grows. Large
stochastic systems often tend to have macroscopic organization due to
limit theorems such as the law of large numbers. This allows us to
identify typical behavior. Our main result is a law of large numbers
for the default rate in the pool; this describes the macroscopically
typical profile. The limiting default rate satisfies an integral
equation that makes explicit the role of the contagion exposure for the
behavior of default clustering in the pool. The result depends heavily
on the analysis of Markov processes via the martingale problem; see
Ethier and Kurtz~\cite{MR88a60130}. We will have more to say on
the mathematical aspects of this in a moment. Once the typical behavior
has been identified, one can then study Gaussian fluctuations and the
structure of atypically large default clusters in the portfolio. We
plan to pursue these directions in a future work.

Previous studies have analyzed the behavior of defaults in large
portfolios. Dembo, Deuschel and Duffie~\cite{MR2022976} examine
a doubly-stochastic model of default timing. In their model, default
correlation is due to the exposure of firms to a common systematic risk
factor which is represented by a random variable. Conditional on this
variable, defaults are independent. A large deviation argument leads to
an approximation of the tail of the conditional portfolio loss
distribution. Glasserman, Kang and Shahabuddin~\cite{Glasserman}
study a copula model of default timing using large deviation
techniques. In that formulation, default events are conditionally
independent given a set of common risk factors. Bush, Hambly, Haworth,
Jin and Reisinger~\cite{hambly} prove a law of large numbers for
a related dynamic model. Davis and Rodriguez~\cite{davis-rod}
develop a law of large numbers and a central limit theorem for the
default rate in a stochastic network setting, in which firms default
independently of one another conditional on the realization of a
systematic factor governed by a finite state Markov chain. Sircar and
Zariphopoulou~\cite{sircar-zari} examine large portfolio
asymptotics for utility indifference valuation of securities exposed to
the losses in the pool. As with these papers, our model includes
exposure to a common systematic risk factor. In contrast, however, our
model captures the self-exciting nature of defaults. Therefore, the
firms in the pool are correlated even after conditioning on the path of
the systematic factor process.\looseness=-1

The use of interacting particle systems to study the behavior of
default clustering in large portfolios is a growing area. In a
mean-field model, Dai Pra, Runggaldier, Sartori and Tolotti
\cite
{daipra-etal} and Dai Pra and Tolotti~\cite{daipra-tolotti} take
the intensity of a constituent firm as a deterministic function of the
percentage portfolio loss due to defaults. In a model with local
interaction, Giesecke and Weber~\cite{giesecke-weber} take the
intensity of a constituent firm as a deterministic function of the
state of the firms in a specified neighborhood of that firm. The
interacting particle perspective leads to the study of the convergence
of interacting Markov processes, laws of large numbers for the
percentage portfolio loss, and Gaussian approximations to the portfolio
loss distribution based on central limit theorems. The interacting
particle system which we propose and study incorporates an additional
source of clustering, namely, the exposure of a firm to a systematic
risk factor process. Moreover, firm-specific sources of default risk
are present in our system. Also, the nature of mean-field interaction
in our system is different. In~\cite{daipra-etal} and \cite
{daipra-tolotti}, a constituent intensity is a function of the current
default rate in the pool. In that formulation, the impact of a default
on the dynamics of the surviving firms is permanent. In our work, a
constituent intensity depends on the path of the default rate. The
impact of a default on the surviving firms is transient, and fades away
exponentially with time. There is a recovery effect.

As we were finishing this work, we learned of a related law of large
numbers type result by Cvitani{\'c}, Ma and Zhang~\cite{CMZ}.
They take the intensity of a constituent firm as a function of a
firm-specific risk factor, a systematic risk factor and the percentage
portfolio loss due to defaults. The risk factors follow diffusion
processes whose coefficients may depend on the portfolio loss. Our
model of the risk factors is more specific than theirs, and thus we are
able to arrive at slightly more explicit results. Moreover, the effect
of defaults in~\cite{CMZ} is permanent, as in~\cite{daipra-etal} and
\cite{daipra-tolotti}.

There are several mathematical contributions in our efforts. Our
analysis of typical events (a weak convergence result) is somewhat
similar to that of certain genetic models (most notably the
Fleming--Viot process; see Chapter~10 of~\cite{MR88a60130}, Fleming and
Viot~\cite{FlemingViot} and Dawson and Hochberg \cite
{DawsonHochberg}),
but the specific form of\vadjust{\goodbreak} our intensity processes imply both
complications and simplifications. Our work is centered on a
jump-diffusion intensity process which is driven by Ornstein--Uhlenbeck
and square root diffusion terms. This formulation allows some explicit
simplifying calculations which are not available in a more abstract
framework. On the other hand, due to the square root singularity,
certain technical estimates need to be developed from scratch (see
Section~\ref{SAppendix}). A final point of interest is heterogeneity.
Interacting particle systems are often assumed to have homogeneous
dynamics, where various parameters are the same for each particle. This
allows the main mathematical arguments to take their simplest form.
Practitioners in credit risk, however, in reality face an extra problem
in data aggregation, where each firm in a portfolio has its own
statistical parameters. We have framed our weak convergence result to
allow for a distribution of ``types,'' that is, a frequency count of
the different model parameters. This leads us to the correct effective
dynamics of the portfolio and, in particular, to a precise formulation
of the effects of self-excitation (see Remark~\ref{Reffcont}).

The rest of this paper is organized as follows. Section~\ref{SModel}
formulates our model of default timing. We establish that our model is
well-posed via the results of Section~\ref{SWellposednessProperties}.
In Section~\ref{STypical} we identify the limit as the number of firms
in the portfolio goes to infinity---a law of large numbers result. The
proof of this result is in Section~\ref{SMainProof}, but depends upon
the technical calculations of Sections~\ref{SLimitIdentification},
\ref
{STightness} and~\ref{SUniqueness}. Section~\ref{Sconclusion}
concludes and discusses extensions. Section~\ref{SAppendix} contains a
number of technical results on square-root-like processes which are
used in our calculations.

\section{Model, assumptions and notation}\label{SModel}
We construct a point process model of correlated default timing in a
portfolio of firms. We assume that $(\Omega,\filt,\BP)$ is an
underlying probability triple on which all random variables are defined.
Let $\{W^n\}_{n\in\N}$ be a countable collection of standard Brownian
motions. Let $\{\ee_n\}_{n\in\N}$ be an i.i.d. collection of standard
exponential random variables. Finally, let $V$ be a standard Brownian
motion which is independent of the $W^n$'s and $\ee_n$'s. Each $W^n$
will represent a source of risk which is idiosyncratic to a specific
firm. Each $\ee_n$ will represent a normalized default time for a
specific firm. The process $V$ will drive a systematic risk factor
process to which all firms are exposed.

Fix an $N\in\N$, $n\in\{1,2,\ldots, N\}$ and consider the following system:
%
\begin{eqnarray} \label{Emain}
d\lambda^\NN_t & = &-\alpha_\NN(\lambda^\NN_t-\bar\lambda_\NN
)\,dt +
\sigma_\NN\sqrt{\lambda^\NN_t}\,dW^n_t \nonumber\\
& &{} + \beta^C_\NN \,dL^N_t+ \eps_N\beta^S_\NN\lambda^\NN_t
\,dX_t,\qquad
t>0, \nonumber\\
\lambda^\NN_0 &= & \lambda_{\circ, N,n}, \nonumber\\[-8pt]\\[-8pt]
dX_t &= & -\gamma X_t \,dt + dV_t,\qquad t>0, \nonumber\\
X_0&= & x_\circ,\nonumber\\
L^N_t &= & \frac{1}{N}\sum_{n=1}^N \chi_{[\ee_n,\infty)}
\biggl(\int
_{s=0}^t \lambda^\NN_s \,ds\biggr). \nonumber
\end{eqnarray}
Here, $\beta^C_\NN\in\R_+\Def[0,\infty)$ and $\beta^S_\NN\in\R $ are
constants which represent the exposure of the $n$th firm in the pool to
$L^N$ and $X$, respectively. The $\alpha_\NN$'s, $\bar\lambda_\NN$'s
and $\sigma_\NN$'s are in $\R_+$ and characterize the dynamics of the
firms. We will address the role of $\eps_N$ in a moment. The initial
condition $x_\circ $ of $X$ is fixed and $\gamma>0$. We use $\chi$ to
represent the indicator function here and throughout the paper.
The\vspace*{1pt} description of $L^N$ is equivalent to a more standard
construction. In particular, define
%
\begin{equation}\label{Etau}
\tau^\NN\Def\inf\biggl\{ t\ge0\dvtx \int_{s=0}^t \lambda^\NN_s \,ds\ge\ee
_n\biggr\}.
\end{equation}
Then
%
\begin{equation}\label{indic} \chi_{[\ee_n,\infty)}\biggl(\int_{s=0}^t
\lambda^\NN_s \,ds\biggr) = \chi_{\{\tau^\NN\le t\}}
\end{equation}
and, consequently,
\[
L^N_t =\frac{1}{N}\sum_{n=1}^N \chi_{\{\tau^\NN\le t\}}.
\]

The process $\lambda^\NN$ represents the intensity, or conditional
event rate, of the $n$th firm in a portfolio of $N$ firms. More
precisely, $\lambda^\NN$ is the instantaneous Doob--Meyer compensator to
the default indicator process (\ref{indic}); see (\ref{EDoobMeyer}).
We will see in Proposition~\ref{Pwellposed} in Section \ref
{SWellposednessProperties} that the $\lambda^\NN$'s are indeed
nonnegative. The process $X$ represents a source of systematic risk;
in our model
this is a stable Ornstein--Uhlenbeck process. The process $L^N$ is the
default rate in the pool.
The jump-diffusion model for $\lambda^\NN$ captures several sources of
default clustering. A firm's intensity is driven by an idiosyncratic
source of risk represented by a Brownian motion $W^n$, and a source of
systematic risk common to all firms---the process $X$. Movements in $X$
cause correlated changes in firms' intensities and thus provide
a source of default clustering emphasized by~\cite{ddk} for corporate
defaults in the U.S. The sensitivity of $\lambda^\NN$ to changes in $X$
is measured by the parameter $\beta^S_\NN$. The second source of
default clustering is through the feedback (``contagion'') term $\beta
^C_\NN \,dL^N_t$. A~default causes an upward jump of size $\frac1N\beta
^C_\NN$ in the intensity $\lambda^\NN$. Due to the mean-reversion of
$\lambda^\NN$, the impact of a default fades away with time,
exponentially with rate $\alpha_\NN$. Self-exciting effects of this
type have been found to be an important source of the clustering of
defaults in the U.S., over and above any clustering caused by the
exposure of firms to systematic risk factors \cite
{azizpour-giesecke-schwenkler}.\vadjust{\goodbreak}

In the special case that $\beta^C_\NN=\beta^S_\NN=0$ for all $n\in
\{
1,2,\ldots, N\}$, the intensities $\lambda^\NN$ follow independent
square root processes so firms default independently of one another.
The formulation (\ref{Emain}) is a natural generalization of the
widely used square root model to address the clustering between defaults.

The interest in large pools of assets is that they provide
\textit{diversification}; they allow one to construct portfolios which have
small variance. The dynamics of $X$ imply that $X$ is stochastically of
order $1$, that is, it is stable.\footnote{Regulatory agencies, for
example, are charged with preventing systematic factors from spiraling
out of control.} Thus, the only way for the pool to have small
variance in our model is for each of the constituent firms to have
small exposure to $X$. We thus assume that
\[
\lim_{N\to\infty}\eps_N=0.
\]
If $\eps_N$ is not small, the influence of the systematic risk factor
$X$ will
be of order~1, and the ``typical'' behavior of the pool will strongly
depend on $X$ (and the tail
behavior of the whole system will be strongly determined by the tail of $X$).
%
\begin{remark} Given the simple structure of $X$, our model is
equivalent, if $x_\circ=0$, to a model where each intensity has
exposure of order $1$ to a small systematic risk. Namely, if $x_\circ=0$,
then $\eps_N X = \tilde X^N$ where
\[
d\tilde X^N_t = -\gamma\tilde X^N_t \,dt + \eps_N \,dV_t.
\]
\end{remark}

Our model allows for a significant amount of bottom-up heterogeneity;
the intensity dynamics of each firm can be different. We capture these
different dynamics by defining the ``types''
%
\begin{equation}\label{Etypedef} \mathsf{p}^\NN\Def(\alpha_\NN
,\bar\lambda
_\NN,\sigma_\NN,\beta^C_\NN,\beta^S_\NN);
\end{equation}
the $\mathsf{p}^\NN$'s take values in parameter space $\PP\Def\R
_+^4\times\R
$. In order to expect regular macroscopic behavior of $L^N$ as $N\to
\infty$, the $\mathsf{p}^\NN$'s and
the $\lambda_{\circ,N,n}$'s should have enough regularity as $N\to
\infty$.
For each $N\in\N$, define
\[
\pi^N \Def\frac{1}{N}\sum_{n=1}^N \delta_{\mathsf{p}^\NN}
\quad\mbox
{and}\quad \Lambda^N_\circ\Def\frac{1}{N}\sum_{n=1}^N \delta
_{\lambda
_{\circ,N,n}};
\]
these are elements of $\Pspace(\PP)$ and $\Pspace(\R_+)$,
respectively.\footnote{As usual, if $E$ is a topological space,
$\Pspace
(E)$ is the collection of Borel probability measures on $E$.}

We need two main assumptions. First, we assume that the types of (\ref
{Etypedef}) and the initial distributions (the $\lambda_{\circ,N,n}$'s)
are sufficiently regular.
%
\begin{assumption}\label{Aregularity} We assume that
\[
\pi\Def\lim
_{N\to\infty}\pi^N\quad \mbox{and}\quad \Lambda_\circ\Def\lim_{N\to\infty
}\Lambda
^N_\circ
\]
exist [in $\Pspace(\PP)$ and $\Pspace(\R_+)$,
resp.].
\end{assumption}

Note that this is what happens in practice; one constructs a
frequency count of the parameters of the different assets in a large
pool and uses this to seek aggregate dynamics for the pool itself. For
a large pool, one hopes that this frequency count will have some
simpler macroscopic description. Second, we assume that the types are bounded.
%
\begin{assumption}\label{ABounded} We assume that there is a $\KK
_{\mmbox{\ref
{ABounded}}}>0$ such that the $\alpha_\NN$'s, $\bar\lambda_\NN$'s,
$\sigma_\NN$'s,
$|\beta^C_\NN|$'s, $|\beta^S_\NN|$'s and $\lambda_{\circ,N,n}$'s are
all bounded by $\KK_{\mmbox{\ref{ABounded}}}$ for all $N\in\N$ and $n\in\{
1,2,\ldots, N\}$.
\end{assumption}

Equivalently, we require that the $\pi_N$'s and $\Lambda
^N_\circ$'s all (uniformly in $N$) have compact support. We could relax
this requirement, at the cost of a much more careful error analysis.

We are interested in the \textit{typical behavior} of $\{L^N\}$. In
Section~\ref{SWellposednessProperties} we consider the well-posedness
of the model (\ref{Emain}), while in Section~\ref{STypical} we state
the law of large numbers result, Theorem~\ref{TMainLLN}.

\section{Well-posedness of the model}\label{SWellposednessProperties}

We here state several technical results concerning the intensities
which are a central part of our model. We want to understand the
structure of the $\lambda^\NN$'s a bit more.
The complications which require
our attention are the square root singularity, and the fact that the
$\lambda_t \,dX_t$ term contains the term $\lambda_t X_t \,dt$, implying
that the dynamics of the $\R^2$-valued
process $(\lambda,X)$ contain a superlinear drift. The proofs of the
results here will be given in Section~\ref{SAppendix}.

Let $W^*$ be a reference Brownian motion with respect to a filtration
$\{\gilt_t\}_{t\ge0}$. Assume also that $V$ is adapted to $\{\gilt _t\}
_{t\ge0}$. Let $\xi$ be a $\{\gilt_t\}_{t\ge0}$-adapted, point process which takes
values in $[0,1]$ and such that $\xi_0=0$. Fix
$\mathsf{p}=(\alpha,\bar \lambda,\sigma ,\beta^C,\beta^S)\in\PP$ and
$\lambda_\circ$ in $\R_+$. Consider the SDE
%
\begin{eqnarray} \label{ElambdaSDE} d\lambda_t &=& -\alpha(\lambda
_t-\bar\lambda)\,dt + \sigma
\sqrt{\lambda_t\vee0}\,dW^*_t +\beta^C \,d\xi_t + \beta^S \lambda_t
\,dX_t,\qquad
t>0,\hspace*{-28pt}\nonumber\\[-8pt]\\[-8pt]
\lambda_0 &=& \lambda_\circ.
\nonumber
\end{eqnarray}
Note that by expanding the dynamics of $dX$ and rearranging a bit, we
get that
\[
d\lambda_t =-\{\alpha+\beta^S\gamma X_t \}\lambda_t \,dt + \alpha
\bar\lambda \,dt + \beta^C \,d\xi_t + \sigma\sqrt{\lambda_t\vee0}\,dW^*_t
+\beta^S \lambda_t \,dV_t.
\]
Also, we have for the moment subsumed the small parameter $\eps_N$ into
the $\beta^S$ term; see the proof of Proposition~\ref{Pwellposed}.

We will use a number of ideas from~\cite{MR1011252} (see also \cite
{Deelsra-Delbaen}).
%
\begin{lemma}\label{Lexistence} There is a nonnegative solution
$\lambda$ of the $\R$-valued SDE (\ref{ElambdaSDE}).
Furthermore, $\sup_{t\in[0,T]}\BE[|\lambda_t|^p]<\infty$ for all $T>0$
and $p\ge1$.
\end{lemma}

We also have uniqueness.
%
\begin{lemma}\label{Luniqueness} The solution of (\ref{ElambdaSDE})
is unique.
\end{lemma}

The model (\ref{Emain}) is thus well posed.
%
\begin{proposition}\label{Pwellposed} The system (\ref{Emain}) has a
unique solution such that $\lambda^{N,n}_t\geq0$ for every $N\in\N$,
$n\in\{1,2,\ldots, N\}$ and $t\geq0$.
\end{proposition}
\begin{pf} Using Lemmas~\ref{Lexistence} and~\ref{Luniqueness}, solve
(\ref{Emain}) between the default times. Replace $\beta^S$ by $\eps_N
\beta^S$ in applying Lemma~\ref{Lexistence}.
\end{pf}

We shall also need a macroscopic bound on the intensities.
%
\begin{lemma}\label{Lmacrobound} For each $p\ge1$ and $T\ge0$,
\[
\KK_{p,T,\mmbox{\ref{Lmacrobound}}} \Deff\mathop{\sup_{0\le t\le T}}_{N\in
\N
}\frac{1}{N}\sum_{n=1}^N\BE[|\lambda^\NN_t|^p]
\]
is finite.
\end{lemma}

\section{Typical events: A law of large numbers}\label{STypical}

Our first task is to understand the ``typical'' behavior of our system.
To do so, we need to understand a system which contains a bit more
information than the default rate $L^N$. For each $N\in\N$ and $n\in
\{
1,2,\ldots, N\}$, define
%
\begin{equation}\label{EDoobMeyer} \dfi^\NN_t \Def\chi_{[0,\ee
_n)}\biggl(\int_{s=0}^t \lambda^\NN_s \,ds\biggr) = \chi_{\{\tau
^\NN>t\}}
\end{equation}
[where $\tau^\NN$ is as in (\ref{Etau})]. In other words, $\dfi
^\NN
_t=1$ if and only if the $n$th firm is still alive at time $t$;
otherwise $\dfi^\NN_t=0$.
Thus, $\dfi^\NN$ is nonincreasing and right-continuous.
It is easy to see that
\[
\dfi^\NN_t + \int_{s=0}^t \lambda^\NN_s \dfi^\NN_s \,ds
\]
is a martingale.
Define $\hat\PP\Def\PP\times\R_+$. For each $N\in\N$, define
$\hat
\mathsf{p}^\NN_t \Def(\mathsf{p}^\NN,\lambda^\NN_t)$
for all $n\in\{1,2,\ldots, N\}$ and $t\ge1$. For each $t\ge0$, define
\[
\mu^N_t \Def\frac{1}{N}\sum_{n=1}^N\delta_{\hat\mathsf{p}^\NN
_t}\dfi^\NN_t;
\]
in other words, we keep track of the empirical distribution of the type
and credit spread for those assets which are still ``alive.''
We note that
\[
L^N_t =1-\mu^N_t(\hat\PP)
\]
for all $t\ge0$.

We want to understand the dynamics of $\mu^N_t$ for large $N$ (this
will then imply the ``typical'' behavior for $L^N_t$).
To understand what our main result is, let's first set up a topological
framework to understand convergence of $\mu^N$. Let $E$ be the
collection of sub-probability measures (i.e., defective probability
measures) on $\hat\PP$, that is, $E$ consists
of those Borel measures $\nu$ on $\hat\PP$ such that $\nu(\hat\PP
)\le1$.
We can topologize $E$ in the usual way (by projecting onto the
one-point compactification of $\hat\PP$; see
\cite{MR90g00004}, Chapter 9.5). In particular, fix a point $\pt$ that is not in
$\hat\PP$ and define $\hat\PP^+\Def\hat\PP\cup\{\pt\}$.
Give $\hat\PP^+$ the standard topology; open sets are those which are
open subsets of $\hat\PP$ (with its original topology) or complements
in $\hat\PP^+$ of closed subsets
of $\hat\PP$ (again, in the original topology of $\hat\PP$). Define a
bijection $\iota$ from $E$ to $\Pspace(\hat\PP^+)$ (the collection of
Borel probability measures
on $\hat\PP^+$) by setting
\[
(\iota\nu)(A) \Def\nu(A\cap\hat\PP) + \bigl(1-\nu(\hat\PP
)
\bigr)\delta_{\pt}(A)
\]
for all $A\in\Borel(\hat\PP^+)$. We can define the Skorohod topology
on $\Pspace(\hat\PP^+)$, and define a corresponding metric on $E$ by
requiring $\iota$ to be an isometry. This makes $E$ a Polish space.
We thus have that $\mu^N$ is an element\footnote{If $S$ is a Polish
space, then $D_S[0,\infty)$ is the collection of maps from $[0,\infty)$
into $S$ which
are right-continuous and which have left-hand limits. The space
$D_S[0,\infty)$ can be topologized by the Skorohod metric, which we
will denote by $d_S$; see Chapter 3.5 of~\cite{MR88a60130}.} of
$D_E[0,\infty)$.

The main theorem of this section is Theorem~\ref{TMainLLN},
essentially a law of large numbers. The construction of the limiting
process will take several steps.
First, for each $\mathsf{p}=(\alpha,\bar\lambda,\sigma,\beta
^C,\beta^S)\in\PP
$, let $b^\mathsf{p}$ satisfy
%
\begin{eqnarray}\label{EDuffiePanSingleton}
\dot b^\mathsf{p}(t) &=& 1-\tfrac12 \sigma^2 (b^\mathsf
{p}(t))^2 - \alpha
b^\mathsf{p}(t) ,\qquad t>0,\nonumber\\[-8pt]\\[-8pt]
b^\mathsf{p}(0)&=&0.
\nonumber
\end{eqnarray}
Note that if $b^\mathsf{p}(t)=0$, then
$\dot b^\mathsf{p}(t)=1>0$. Thus, $b^\mathsf{p}(t)>0$ for all $t>0$.

The next lemma is essential for the characterization of the limit.
Its proof is deferred to Section~\ref{SAppendix}.
%
\begin{lemma}\label{LbQDef} There is a unique $\R_+$-valued trajectory
$\{Q(t); t\ge0\}$
which satisfies the equation
%
\begin{eqnarray}\label{EbQDef} Q(t) &=& \mathop{\int_{\hat\mathsf
{p}=(\mathsf{p},\lambda)\in\hat
\PP}}_{
\mathsf{p}=(\alpha,\bar\lambda,\sigma,\beta^C,\beta^S)} \beta^C
\biggl[ \dot
b^\mathsf{p}(t)\lambda+ \int_{r=0}^t\dot b^\mathsf{p}(t-r)\{ Q(r)+
\alpha\bar
\lambda\}\,dr\biggr] \nonumber\\
&&\hspace*{61pt}{}\times\exp\biggl[-b^\mathsf{p}(t)\lambda- \int_{r=0}^t
b^\mathsf{p}(t-r)\{ Q(r) + \alpha\bar\lambda\} \,dr
\biggr]\\
&&\hspace*{61pt}{}\times\pi(d\mathsf{p})\Lambda_\circ(d\lambda).
\nonumber
\end{eqnarray}
Here, $\pi$ and $\Lambda_\circ$ are as in Assumption~\ref{Aregularity}.
\end{lemma}

Now let $W^*$ be a reference Brownian motion. For each $\hat
\mathsf{p}=(\mathsf{p},\lambda_\circ)\in
\hat\PP$ where $\mathsf{p}= (\alpha,\bar
\lambda,\sigma,\beta^C,\beta^S)$, let $\lambda^*_t(\mathsf{p})$
be the
unique solution to
%
\begin{eqnarray}\label{EEffectiveEquation1}
\lambda^*_t(\hat\mathsf{p}) &=& \lambda_\circ-
\alpha\int
_{s=0}^t \bigl(\lambda^*_s(\hat\mathsf{p})-\bar\lambda\bigr)\,ds + \sigma\int
_{s=0}^t\sqrt{\lambda^*_s(\hat\mathsf{p})}\,dW^*_s
\nonumber\\[-8pt]\\[-8pt]
&&{}+ \int_{s=0}^t
Q(s) \,ds.\nonumber
\end{eqnarray}
We now have our main result.
%
\begin{theorem}\label{TMainLLN} For all $A\in\Borel(\PP)$ and
$B\in
\Borel(\R_+)$, define
%
\begin{eqnarray}\label{Emudef} \bar\mu_t(A\times B)
&\Deff&\int
_{\hat\mathsf{p}= (\mathsf{p},\lambda)\in\hat\PP} \chi
_A(\mathsf{p}) \BE\biggl[\chi
_B(\lambda^*_t(\hat\mathsf{p}))\exp\biggl[-\int_{s=0}^t \lambda
_s^*(\hat\mathsf{p}
)\,ds\biggr]\biggr]\nonumber\\[-8pt]\\[-8pt]
&&\hspace*{45pt}{}\times
\pi(d\mathsf{p})\Lambda_\circ(d\lambda).\nonumber
\end{eqnarray}
Then
%
\begin{equation}\label{Emulimit} \lim_{N\to\infty}\BP\bigl\{
d_{\Pspace
(\hat\PP)}(\mu^N,\bar\mu)\ge\delta\bigr\}= 0
\end{equation}
for every $\delta>0$. Define
%
\begin{eqnarray}\label{EFrep}
&&F(t) \Deff1-\bar\mu_t(\hat\PP) \nonumber\\[-8pt]\\[-8pt]
&&\qquad= 1-\int_{\hat\mathsf{p}= (\mathsf{p},\lambda)\in\hat\PP} \BE
\biggl[\exp\biggl[-\int
_{s=0}^t \lambda_s^*(\hat\mathsf{p})\,ds\biggr]\biggr] \pi(d\mathsf
{p})\Lambda_\circ
(d\lambda). \nonumber
\end{eqnarray}
Then, for all $\delta>0$ and $T>0$,
\[
\lim_{N\to\infty}\BP\Bigl\{\sup_{0\le t\le
T}|L^N_t-F(t)|\ge\delta\Bigr\}= 0.
\]
\end{theorem}

The parts of the proof of this result will be given in
Sections~\ref{SLimitIdentification},~\ref{STightness} and \ref
{SUniqueness}. In particular, in Section~\ref{SLimitIdentification}
we identify a candidate limit for $\{\mu^{N}\}$ using the martingale
problem formulation. Then in Section~\ref{STightness} we prove that
$\{
\mu^{N}\}$ is tight, which ensures that the laws of $\{\mu^{N}\}$'s
have at least one limit point. In Section~\ref{SUniqueness} we prove
that the limit is necessarily unique. Then, in Section~\ref{SMainProof}
we collect things together to prove Theorem~\ref{TMainLLN}.

With this result in hand, we can rewrite (\ref{EEffectiveEquation1})
to see the exposure of a typical firm to the contagion factor.
%
\begin{remark}\label{RAlternativeRepresentationForLimit}
We have that
%
\begin{eqnarray}\label{EFDer}
\dot F(t) &=& \int_{\hat\mathsf{p}= (\mathsf{p},\lambda)\in\hat
\PP} \BE
\biggl[\lambda_t^*(\hat\mathsf{p})\exp\biggl[-\int_{s=0}^t \lambda
_s^*(\hat\mathsf{p}
)\,ds\biggr]\biggr] \pi(d\mathsf{p})\Lambda_\circ(d\lambda) \nonumber\\[-8pt]\\[-8pt]
& = &\int_{\hat\mathsf{p}= (\mathsf{p},\lambda)\in\hat\PP}
\lambda\bar\mu_t(d\hat
\mathsf{p}).
\nonumber
\end{eqnarray}
Thus,
%
\begin{eqnarray}\label{EEffectiveEquation2}
\lambda^*_t(\hat\mathsf{p}) &=& \lambda_\circ- \alpha\int
_{s=0}^t\bigl(\lambda
^*_s(\hat\mathsf{p})-\bar\lambda\bigr)\,ds + \sigma\int_{0}^t\sqrt
{\lambda^*_s(\hat
\mathsf{p})}\,dW^*_s \nonumber\\[-8pt]\\[-8pt]
&&{}+ \int_{0}^tB(\bar\mu_s) \dot F(s)\,ds,\nonumber
\end{eqnarray}
where
\[
B(\mu) \Def\mathop{\int_{\hat\mathsf{p}=(\mathsf{p},\lambda)\in
\hat\PP}}_{ \mathsf{p}
=(\alpha,\bar\lambda,\sigma,\beta^C,\beta^S)}\beta^C\lambda\mu
(d\hat
\mathsf{p})\Big/\mathop{\int_{\hat\mathsf{p}=(\mathsf
{p},\lambda)\in\hat\PP}}_{ \mathsf{p}
=(\alpha,\bar\lambda,\sigma,\beta^C,\beta^S)} \lambda\mu(d\hat
\mathsf{p})
\]
for all $\mu\in E$. In other words, the effective sensitivity of a
typical intensity to the contagion is given by an average weighted by
the instantaneous intensities. Note that
$0\le B(\mu)\le\KK_{\mmbox{\ref{ABounded}}}$.
\end{remark}

The homogeneous case provides more explicit insights into the role of
the contagion exposure for the behavior of default clustering in the pool.
%
\begin{remark}\label{RLimitInHomogeneousCase} Fix $\hat\mathsf
{p}=(\mathsf{p}
,\lambda_\circ)\in\hat\PP$ where $\mathsf{p}= (\alpha,\bar
\lambda,\sigma
,\beta^C,\beta^S)$. Assume that the pool is homogeneous, that
is,
$\hat
\mathsf{p}^\NN=\hat\mathsf{p}$ for all $N\in\N$ and $n\in\{
1,2,\ldots, N\}$. By the
relation (\ref{EEffectiveEquation2}), we then have that $Q(t)=\beta
^C\dot F(t)$.
In this case, $F$ is given by the unique solution to the integral equation
\[
F(t)=1-\exp\biggl[-\alpha\bar{\lambda}\int_{r=0}^tb^\mathsf
{p}(t-r)\,dr -\beta^C
\int_{r=0}^t F(r)\dot{b}^\mathsf{p}(t-r)\,dr-b^\mathsf{p}(t)\lambda
_\circ\biggr].
\]
Furthermore, if there is no exposure to contagion, that is, $\beta
^C=0$, then this integral equation reduces to the well-known explicit formula
\[
F(t)=1-\exp\biggl[-\alpha\bar{\lambda}\int_{r=0}^tb^\mathsf
{p}(t-r)\,dr-b^\mathsf{p}
(t)\lambda_\circ\biggr].
\]
Figure~\ref{fig1} shows the limiting default rate $F(t)$ for different
values of the contagion sensitivity $\beta^C$. The default rate
increases with $\beta^C$. Figure~\ref{fig2} shows the limiting default
rate $F(t)$ for different values of the parameter $\alpha$, which
%
\begin{figure}

\includegraphics{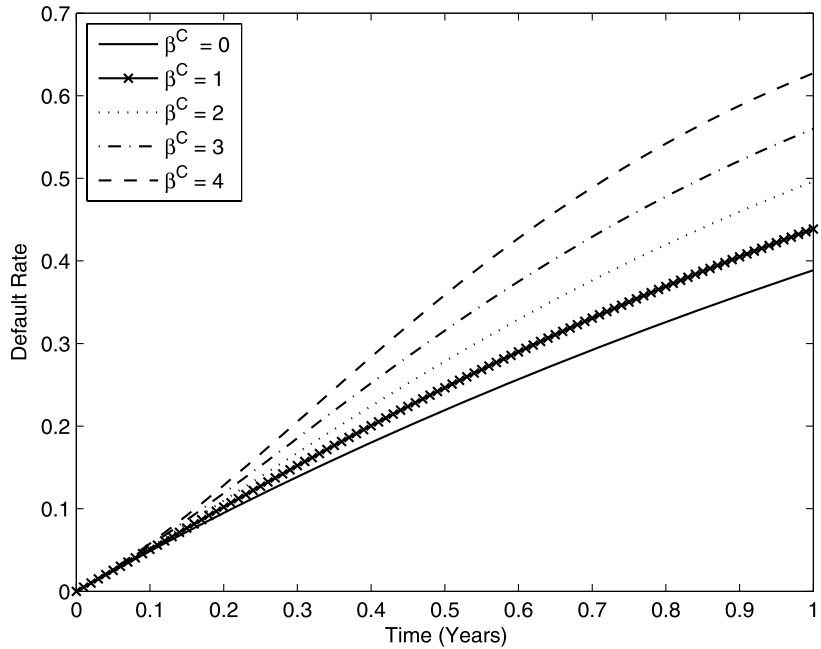}
 \caption{Comparison of limiting default rate $F(t)$
for different values of the contagion sensitivity~$\beta^C$. The
parameter case is $\sigma= 0.9$, $\alpha= 4$, $\bar{\lambda} = 0.5$ and
$\lambda _0 = 0.5$.}\label{fig1}
\end{figure}
specifies the reversion speed of the intensity. The default rate is
relatively insensitive to changes in $\alpha$ for shorter horizons; for
longer horizons it decreases with $\alpha$. The limiting default rate
%
\begin{figure}

\includegraphics{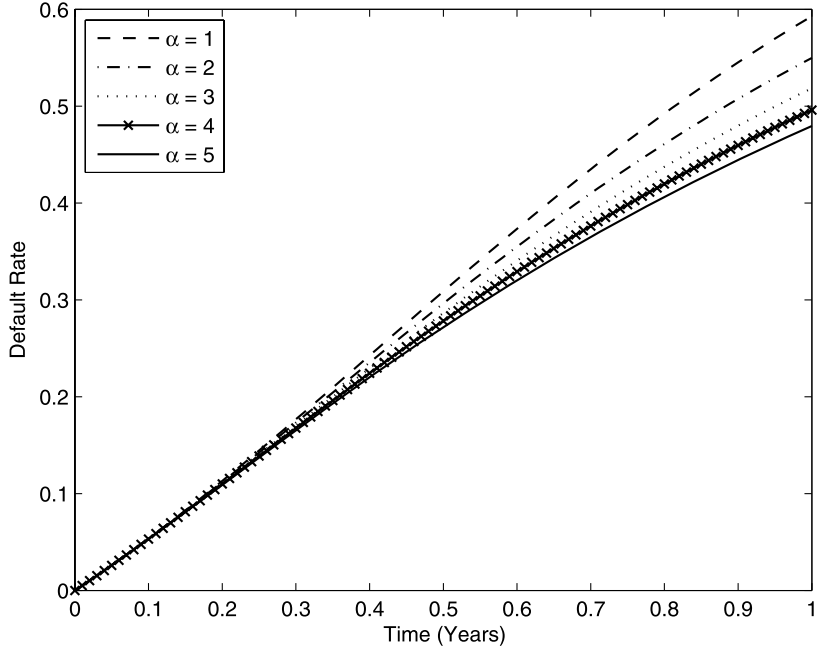}

\caption{Comparison of limiting default rate $F(t)$ for
different values of the reversion speed $\alpha$. The parameter case is
$\sigma= 0.9$, $\beta^C=2$, $\bar{\lambda} = 0.5$ and $\lambda_0 = 0.5$.}
\label{fig2}
\end{figure}
is more sensitive to variation in the reversion level $\bar\lambda$, as
indicated in Figure~\ref{fig3}. Variations in the diffusive volatility
$\sigma$ of the intensity have little effect on $F(t)$.
\end{remark}

\begin{figure}

\includegraphics{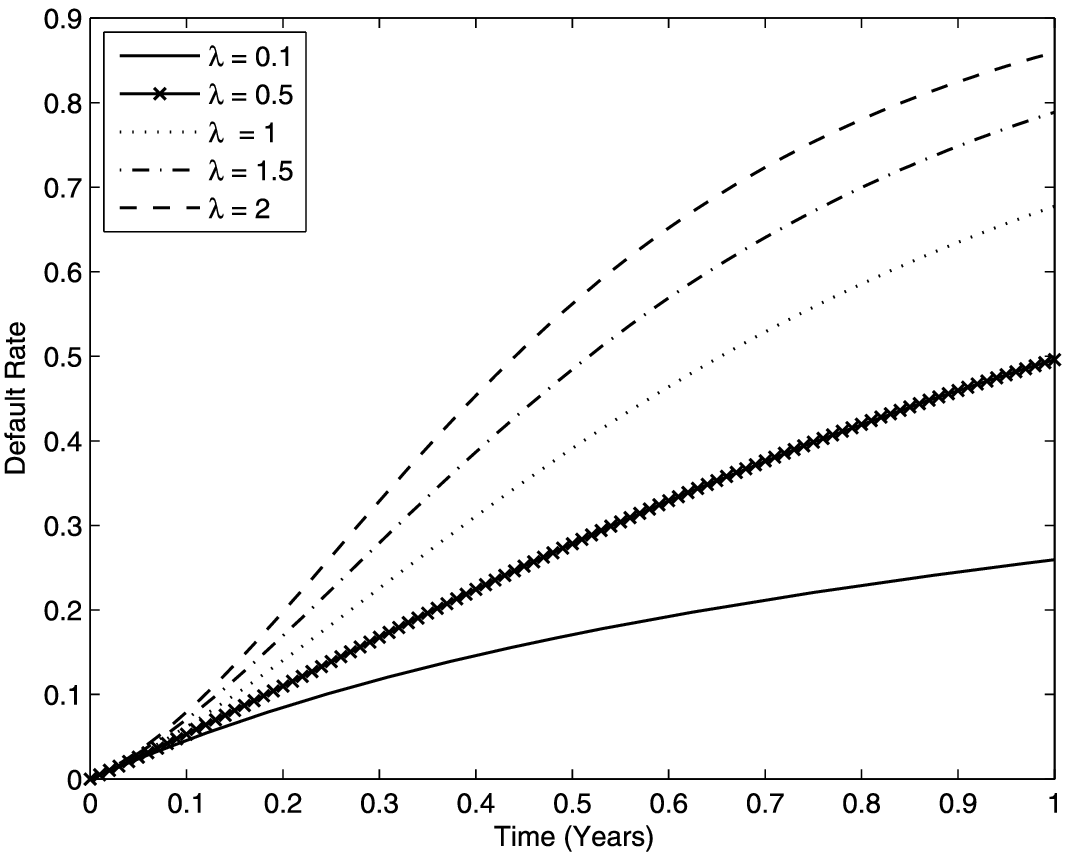}

\caption{Comparison of limiting default rate $F(t)$ for
different values of the reversion level $\bar\lambda$. The parameter
case is $\sigma= 0.9$, $\beta^C=2$, $\alpha= 4$ and $\lambda_0 = 0.5$.}
\label{fig3}
\end{figure}

We finally note that the structure of the unperturbed (i.e., \mbox{$\beta
^C=\beta^S=0$}) dynamics of the intensity (\ref{Emain}) was crucial in
singling out the equation (\ref{EbQDef}) as the proper macroscopic
effect of the contagion (see the proof of Lemma~\ref{LQchar}). The
calculations in fact hinge upon the explicit formulae for affine jump
diffusions developed in~\cite{DuffiePanSingleton}. In a more general
setting we would need a more abstract framework (see~\cite{CMZ}).

\section{Identification of the limit}\label{SLimitIdentification}
We want to use the martingale problem (see Chapter 4 of \cite
{MR88a60130}) to show that $\mu^N$'s converge to a limiting process.
For every $f\in C^\infty(\hat\PP)$ and $\mu\in E$, define
\[
\langle f,\mu\rangle_E \Def\int_{\hat\mathsf{p}\in\hat\PP
}f(\hat\mathsf{p})\mu(d\hat\mathsf{p}
).
\]
Let $\SSS$ be the collection of elements $\Phi$ in $B(\Pspace(\hat
\PP
))$ of the form
%
\begin{equation}\label{Eform} \Phi(\mu) = \varphi(\langle
f_1,\mu\rangle
_E,\langle f_2,\mu\rangle_E,\ldots,\langle f_M,\mu\rangle_E)
\end{equation}
for some $M\in\N$, some $\varphi\in C^\infty(\R^M)$ and some $\{
f_m\}
_{m=1}^M$. Then $\SSS$ separates $\Pspace(\hat\PP)$ (see Chapter 3.4
of~\cite{MR88a60130}). It thus suffices to show convergence of the martingale
problem for functions of the form (\ref{Eform}).

Let's fix $f\in C^\infty(\hat\PP)$ and understand exactly what happens
to $\langle f,\mu^N\rangle_E$ when one of the firms defaults. Suppose
that the
$n$th firm defaults at time $t$ and that none of the other firms
default at time $t$ (defaults occur simultaneously
with probability zero). Then
\begin{eqnarray*} \langle f,\mu^N_t\rangle_E &=&\frac{1}{N}\mathop
{\sum_{1\le
n'\le N }}_{ n'\not= n} f\biggl(\mathsf{p}^{N,n'},\lambda^{N,n'}_t+
\frac{\beta
^C_\NN}{N}\biggr)\dfi^{N,n'}_t,\\
\langle f,\mu^N_{t-}\rangle_E &=& \frac{1}{N}
f(\mathsf{p}^{N,n'},\lambda^{N,n'}_t)\dfi^{N,n'}_t
+\frac1Nf(\mathsf{p}^\NN,\lambda^\NN_t).
\end{eqnarray*}
Note, furthermore, that the default at time $t$ means that $\int
_{s=0}^t \lambda^\NN_s\,ds=\ee_n$, so $\dfi^\NN_t=0$.
Hence,
%
\begin{equation} \label{Ejjumps} \langle f,\mu^N_t\rangle_E-\langle
f,\mu^N_{t-}\rangle
_E = \jump^f_\NN(t),
\end{equation}
where
\begin{eqnarray*}
\jump^f_\NN(t) &\Def& \frac{1}{N}\sum_{n'=1}^N \biggl\{ f\biggl(\mathsf{p}
^{N,n'},\lambda^{N,n'}_t+ \frac{\beta^C_\NN}{N}\biggr)-f
(\mathsf{p}
^{N,n'},\lambda^{N,n'}_t)\biggr\}\dfi^{N,n'}_t \\
&&{} -\frac1Nf(\mathsf{p}^\NN,\lambda^\NN_t)
\end{eqnarray*}
for all $t\ge0$, $N\in\N$ and $n\in\{1,2,\ldots, N\}$.

We now identify the limiting martingale problem for $\mu^N$.
For $\hat\mathsf{p}=(\mathsf{p},\lambda)$ where $\mathsf
{p}=(\alpha,\bar\lambda,\sigma
,\beta^C,\beta^S)\in\PP$ and $f\in C^\infty(\hat\PP)$, define
the operators
%
\begin{eqnarray}\label{EOperators1} (\genL_1 f)(\hat\mathsf{p}) &=&
\frac12 \sigma^{2}\lambda\,\frac
{\partial^2 f}{\partial\lambda^2}(\hat\mathsf{p}) - \alpha
(\lambda-\bar
\lambda)\,\frac{\partial f}{\partial\lambda}(\hat\mathsf
{p})-\lambda f(\hat\mathsf{p}
),\nonumber\\[-8pt]\\[-8pt]
(\genL_2 f)(\hat\mathsf{p}) &=& \frac{\partial f}{\partial\lambda
}(\hat\mathsf{p}).
\nonumber
\end{eqnarray}
Define also
\[
\QQ(\hat\mathsf{p}) \Def\lambda\beta^C
\]
for $\hat\mathsf{p}=(\mathsf{p},\lambda)$ where $\mathsf
{p}=(\alpha,\bar\lambda,\sigma
,\beta^C,\beta^S)\in\PP$.
The generator $\genL_1$ corresponds to the diffusive part of the
intensity with killing rate $\lambda$, and $\genL_2$ is the macroscopic
effect of contagion on the
surviving intensities at any given time.
For $\Phi\in\SSS$ of the form (\ref{Eform}), define
%
\begin{eqnarray} \label{Elimgen}
&&(\genA\Phi)(\mu) \Def \sum_{m=1}^M \frac{\partial\varphi
}{\partial
x_m}(\langle f_1,\mu\rangle_E,\langle f_2,\mu\rangle_E,\ldots,
\langle f_M,\mu\rangle_E)\nonumber\\[-8pt]\\[-8pt]
&&\hspace*{74pt}{}
\times \{\langle \genL_1f_m,\mu\rangle_E
+ \langle\QQ,\mu\rangle_E \langle\genL_2 f_m,\mu
\rangle_E \}.
\nonumber
\end{eqnarray}
We claim that $\genA$ will be the generator of the limiting martingale problem.
%
\begin{lemma}[(Weak convergence)]\label{Lwconv} For any $\Phi\in\SSS$
and $0\le r_1\le r_2\cdots r_J=s<t<T$ and $\{\psi_j\}_{j=1}^J\subset
B(E)$, we have that
\[
\lim_{N\to\infty}\BE\Biggl[\biggl\{\Phi(\mu^N_t)-\Phi(\mu^N_s)-\int_{r=s}^t
(\genA\Phi)(\mu^N_r)\,dr\biggr\}\prod_{j=1}^J \psi_j(\mu^N_{r_j})\Biggr]=0.
\]
\end{lemma}
\begin{pf}
For $\hat\mathsf{p}=(\mathsf{p},\lambda)$ where $\mathsf
{p}=(\alpha,\bar\lambda,\sigma
,\beta^C,\beta^S)\in\PP$, define
\begin{eqnarray*} (\genL^a f)(\hat\mathsf{p}) &=& \frac12 \sigma
^2\lambda\,\frac
{\partial^2 f}{\partial\lambda^2}(\hat\mathsf{p}) - \alpha
(\lambda-\bar
\lambda)\,\frac{\partial f}{\partial\lambda}(\hat\mathsf{p}),\\
(\genL_x^b f)(\hat\mathsf{p}) &=& \lambda\biggl\{\frac12 \,\frac
{\partial^2
f}{\partial\lambda^2}(\hat\mathsf{p}) -\gamma x \,\frac{\partial
f}{\partial\lambda}(\hat\mathsf{p})\biggr\},\qquad x\in\R.
\end{eqnarray*}
Then
$ \genL^a$ is the generator of the idiosyncratic part of the intensity
and $ \genL^b_x$ is the generator of the systematic risk.

We start by writing that
\[
\Phi(\mu^N_t)=\Phi(\mu^N_0)+\int_{r=0}^t \{ A^{N,1}_r + A^{N,2}_r\}
\,dr +\mart_t,
\]
where $\mart$ is a martingale and
\begin{eqnarray*} A^{N,1}_t &=& \sum_{m=1}^M \frac{\partial\varphi
}{\partial x_m}(\langle f_1,\mu^N_t\rangle_E,\langle f_2,\mu
^N_t\rangle_E,\ldots,\langle
f_M,\mu^N_t\rangle_E)\\
&&\hspace*{16.1pt}{} \times\frac{1}{N}\sum_{n=1}^N \{(\genL^a f_m)(\hat\mathsf
{p}^\NN
_t)+ \eps_N( \genL^b_{X_t}f_m)(\hat\mathsf{p}^\NN_t)\} \dfi
^\NN_t\\
&=& \sum_{m=1}^M \frac{\partial\varphi}{\partial x_m}(\langle
f_1,\mu^N_t\rangle_E,\langle f_2,\mu^N_t\rangle_E,\ldots,\langle
f_M,\mu^N_t\rangle_E
)\\
&&\hspace*{16.1pt}{}\times
\{\langle\genL^a f_m,\mu^N_t\rangle_E
+\eps_N\langle\genL_{X_t}^bf_m,\mu^N_t\rangle_E
\} ,\\
A^{N,2}_t &=& \sum_{n=1}^N \lambda^\NN_t\bigl\{\varphi\bigl(\langle
f_1,\mu
^N_t\rangle_E+\jump^{f_1}_\NN(t),\\
&&\hspace*{53.5pt}\langle f_2,\mu^N_t\rangle_E
 +\jump^{f_2}_\NN(t),\ldots,
\langle f_M,\mu^N_t\rangle
_E+\jump
^{f_M}_\NN(t)\bigr) \\
&&\hspace*{34.3pt}\hspace*{39pt}{} -\varphi(\langle f_1,\mu^N_t\rangle_E,\langle
f_2,\mu^N_t\rangle
_E,\ldots,\langle f_M,\mu^N_t\rangle_E)\bigr\}\dfi^\NN_t.
\end{eqnarray*}

Using Lemma~\ref{Lmacrobound}, it is fairly easy to see that for all
$f\in C^\infty(\hat\PP)$,
\[
\lim_{N\rightarrow\infty} \BE\biggl[\eps_N\int_{r=0}^t
|\langle\genL
_{X_r}^b f,\mu^N_r\rangle_E|\,dr\biggr]=0.
\]

To proceed, let's simplify $\jump^f_\NN$. For each $f\in C^\infty
(\hat
\PP)$, $t\ge0$, $N\in\N$ and $n\in\{1,2,\ldots, N\}$, define
\begin{eqnarray*}
\tilde\jump^f_\NN(t) &\Def& \frac{\beta^C_\NN}{N}\sum_{m=1}^N
\frac
{\partial f}{\partial\lambda}(\hat\mathsf{p}^\NN_t)\dfi
^{N,m}_t-f(\mathsf{p}
^\NN,\lambda^\NN_t) \\
& = &\beta^C_\NN\langle\genL_2 f,\mu^N_t\rangle_E-f(\hat\mathsf{p}^N_t).
\end{eqnarray*}
Then
\[
\biggl|\jump^f_\NN(t)-\frac1N\tilde\jump^f_\NN(t)\biggr|\le
\frac{\KK
_{\mmbox{\ref{ABounded}}}^2}{N^2}\biggl\|\frac{\partial^2 f}{\partial
\lambda
^2}\biggr\|_C,
\]
where $\KK_{\mmbox{\ref{ABounded}}}$ is the constant from Assumption
\ref{ABounded}.

Define $\iota(\hat\mathsf{p})\Def\lambda$ for $\hat\mathsf
{p}=(\mathsf{p},\lambda)\in
\hat\PP$. Setting
%
\begin{eqnarray}\label{Eeffcc}
\tilde A^{N,2}_t &\Def&\sum_{m=1}^M \frac{\partial\varphi}{\partial
x_m}(\langle f_1,\mu^N_t\rangle_E,\langle f_2,\mu^N_t\rangle
_E,\ldots,\langle f_M,\mu
^N_t\rangle_E) \nonumber\\
&&\hspace*{16.3pt}{} \times\frac{1}{N}\sum_{n=1}^N \lambda^\NN_t\tilde\jump
^{f_m}_\NN(t) \dfi^\NN_t \nonumber\\[-8pt]\\[-8pt]
&= &\sum_{m=1}^M \frac{\partial\varphi}{\partial x_m}(\langle
f_1,\mu^N_t\rangle_E,\langle f_2,\mu^N_t\rangle_E,\ldots,\langle
f_M,\mu^N_t\rangle_E) \nonumber\\
&&\hspace*{16.8pt}{} \times\{\langle\QQ,\mu^N_t\rangle_E \langle\genL_2
f_m,\mu^N_t\rangle_E-
\langle\iota f,\mu^N_t\rangle_E\},
\nonumber
\end{eqnarray}
we have that
\[
\lim_{N\to\infty}\BE\biggl[\int_{r=0}^t |A^{N,2}_r-\tilde
A^{N,2}_r|\,dr\biggr]=0.
\]
Collecting things together, we have that
\[
\lim_{N\to\infty}\BE\Biggl[\biggl\{\int_{r=s}^t A^{N,1}_r \,dr + \int_{r=s}^t
A^{N,2}_r \,dr - \int_{r=s}^t (\genA\Phi)(\mu^N_r)\,dr\biggr\}\prod_{j=1}^J
\psi_j(\mu^N_{r_j})\Biggr]=0,
\]
which implies the claim.
\end{pf}

We, in particular, note the macroscopic effect of the contagion.
%
\begin{remark}\label{Reffcont} The key step in quantifying the
coarse-grained effect of contagion was (\ref{Eeffcc}). Namely, we
average the combination of the jump rate and the exposure to contagion
across the pool.\vspace*{-3pt}
\end{remark}

\section{Tightness}\label{STightness}

In this subsection we verify that the family $\{\mu^N\}_{N\in\N}$ is
relatively compact (as a $D_E[0,\infty)$-valued random variable); this
of course is necessary to ensure that the laws of the $\mu^N$'s
have at least one limit point. The complication of course is the
feedback through contagion.
We need to show that the system is unlikely to ``explode'' via
feedback. Our calculations are framed by Theorem 8.6 of Chapter 3 of
\cite{MR88a60130}; we need to show compact containment and regularity
of the $\mu^N$'s.

In particular, compact containment ensures that there is a compact set
$\compact$ such that $\mu^N_t$ will belong to $\compact$ for all $N\in
\N$ and $t\in[0,T]$ with high probability; see Lemma
\ref{LCompactContainment}. Regularity shows, roughly speaking, that
$\mu ^N_t-\mu^N_s$ is bounded in a certain sense by a function of the
time interval $t-s$, that goes to zero as the length of the time
interval goes to zero; see Lemma~\ref{LRegularity}. By Theorem 8.6 of
Chapter\vspace*{1pt} 3 of~\cite{MR88a60130}, these two statements imply relative
compactness of the family $\{\mu^N\}_{N\in\N}$ in $D_E[0,\infty)$; see
Lemma~\ref{LMuMeasureTight}.

Let's first address compact containment.
%
\begin{lemma}\label{LCompactContainment} For each $\eta>0$ and $t\ge
0$, there is a compact subset $\compact$ of $E$ such that
\[
\mathop{\sup_{N\in\N}}_{ 0\le t< T}\BP\{\mu^N_t\notin\compact
\}<\eta.
\]
\end{lemma}

\begin{pf} For each $L>0$, define $K_L\Def[-\KK_{\mmbox{\ref
{ABounded}}},\KK
_{\mmbox{\ref{ABounded}}}]^3\times[0,\KK_{\mmbox{\ref{ABounded}}}]^2\times[0,L]$.
Then $K_L \subset\subset\hat\PP$, and for each $t\ge0$ and $N\in
\N$,
\[
\BE[\mu^N_t(\hat\PP\setminus K_L)] = \frac{1}{N}\sum
_{n=1}^N \BP\{\lambda^\NN_t\ge L\}\le\frac{\KK_{1,T,\mmbox{\ref
{Lmacrobound}}}}{L}.
\]
Here $\KK_{\mmbox{\ref{ABounded}}}$ and $\KK_{1,T,\mmbox{\ref{Lmacrobound}}}$ are
the constants from Assumption~\ref{ABounded} and Lemma~\ref{Lmacrobound}.
Let's next define
\[
\compact^*_L \Def\overline{\biggl\{\nu\in E \dvtx\nu\bigl(\hat\PP
\setminus K_{(L+j)^2}\bigr) < \frac{1}{\sqrt{L+j}}\mbox{ for all }j\in\N\biggr\}};
\]
these are compact subsets of $E$.
We have that
\begin{eqnarray*} \BP\{\mu^N_t\notin\compact^*_L\}&\le&\sum
_{j=1}^\infty\BP\biggl\{\mu^N_t\bigl(\hat\PP\setminus K_{(L+j)^2} \bigr)> \frac
{1}{\sqrt{L+j}}\biggr\}\\[-2pt]
& \le&\sum_{j=1}^\infty\frac{\BE[\mu^N_t(\hat\PP\setminus
K_{(L+j)^2})]}{1/\sqrt{L+j}}\\[-2pt]
&\le&\sum_{j=1}^\infty\frac{\KK_{1,T,\mmbox{\ref
{Lmacrobound}}}}{(L+j)^2/\sqrt{L+j}}
\le\sum_{j=1}^\infty\frac{\KK_{1,T,\mmbox{\ref{Lmacrobound}}}}{(L+j)^{3/2}}.
\end{eqnarray*}
Since
\[
\lim_{L\to\infty}\sum_{j=1}^\infty\frac{\KK_{1,T,\mmbox{\ref
{Lmacrobound}}}}{(L+j)^{3/2}} =0,
\]
the result follows.
\end{pf}

We next need to understand the regularity of the $\mu^N$'s.
For each $t\ge0$ and \mbox{$N\in\N$}, we define
\[
\filt^N_t \Def\sigma\bigl\{\lambda^{\NN}_{s}; 0\le s\le t, n\in\{
1,2,\ldots, N\}\bigr\}.
\]
Let's also define $q(x,y) \Def\min\{|x-y|,1\}$ for all $x$ and $y$ in
$\R$.

To proceed, let's first consider the $L^N$'s. A useful tool will be the
following integral bound. Fix $T>0$ and suppose that $f$ is a
square-integrable function on $[0,T]$.
Then for any $0\le s\le t\le T$,
%
\begin{eqnarray}\label{Eintbnd} \int_{r=s}^t f(r)\,dr &\le&\sqrt
{t-s}\sqrt{\int_{r=0}^T
f^2(r)\,dr} \nonumber\\
& \le&\frac12 \biggl\{\frac{\sqrt{t-s}}{(t-s)^{1/4}} + (t-s)^{1/4}\int
_{r=0}^T f^2(r)\,dr\biggr\}\\
&= &\frac12(t-s)^{1/4}\biggl\{1+ \int_{r=0}^T f^2(r)\,dr\biggr\}.
\nonumber
\end{eqnarray}

\begin{lemma}\label{Ltight} Define
\[
\Xi_N \Deff\frac{1}{2N}\sum_{n=1}^N\biggl\{1 + \int_{r=0}^t
(\lambda
^\NN_r)^2 \,dr\biggr\}= \frac12\Biggl\{1 + \frac{1}{N}\sum_{n=1}^N\int
_{r=0}^T (\lambda^\NN_r)^2 \,dr\Biggr\}.
\]
Then $ \BE[\Xi_N] \le\frac12\{1+ \KK_{2,T,\mmbox{\ref{Lmacrobound}}}\}$
(where $\KK_{2,T,\mmbox{\ref{Lmacrobound}}}$ is the constant from
Lemma~\ref{Lmacrobound}) and
\[
\BE[|L^N_t-L^N_s| |\filt^N_s] \le(t-s)^{1/4}\BE[\Xi_N|\filt^N_s]
\]
for all $0\le s\le t\le T$.
\end{lemma}
\begin{pf}
The bound $ \BE[\Xi_N] \le\frac12\{1+ \KK_{2,T,\mmbox{\ref
{Lmacrobound}}}\}
$ is clear from Lemma~\ref{Lmacrobound}. To proceed, let's write
\[
L^N_t = 1-\frac{1}{N}\sum_{n=1}^N \dfi^\NN_t.
\]
By the martingale problem for $L^N$, we have that $L^N_t=A^N_t+\mart_t$
where $\mart$ is a martingale and where
\[
A^N_t = \frac{1}{N}\sum_{n=1}^N\int_{r=0}^t\lambda^\NN_r \dfi^\NN
_r \,dr.
\]
Thus, for $0\le s\le t$, we have (keeping in mind that $L^N$ is nondecreasing)
\[
|L^N_t-L^N_s| = L^N_t-L^N_s = A^N_t-A^N_s + \mart_t-\mart_s.
\]
We then can use (\ref{Eintbnd}) to see that
\[
A^N_t-A^N_s \le\frac{1}{N}\sum_{n=1}^N\int_{r=0}^t \lambda^\NN_r
dr\le
(t-s)^{1/4}\Xi_N.
\]
The claimed bound follows.
\end{pf}

Of course, $\BP\{L^N_t\in[0,1]\}=1$ for all $t\ge0$ and
$N\in\N$, so compact containment (i.e., condition (a)
of Theorem 7.2 of Chapter 2 of~\cite{MR88a60130}) definitely holds.

Moreover, by Lemma~\ref{Ltight} we have that for any $0\le t\le T$,
$0\le u\le\delta$, and $0\le v\le\delta\wedge t$,
\[
\BE[q(L^N_{t+u},L^N_t)| \filt^N_t] q(L^N_t,L^N_{t-v})
\le\BE[L^N_{t+u}-L^N_t| \filt^N_t] \\
\le\delta^{1/4}\BE[\Xi_N|\filt^N_t].
\]
Theorem 8.6 of Chapter 3 of~\cite{MR88a60130} thus implies that $\{
L^N\}_{N\in\N}$ is relatively compact.
%
\begin{lemma} \label{LRegularity}
There is a random variable $H_{N}$ with $\sup_{N\in\N}\BE
[H_N]<\infty
$, such that for any $0\le t\le T$, $0\le u\le\delta$, and $0\le v\le
\delta\wedge t$,
\[
\BE[q^2(\langle f,\mu^N_{t+u}\rangle_E,\langle f,\mu
^N_t\rangle_E)q^2(\langle f,\mu
^N_t\rangle_E,\langle f,\mu^N_{t-v}\rangle_E)| \filt^N_t
] \le\delta
^{1/4}\BE[H_N| \filt^N_t].
\]
\end{lemma}
\begin{pf}
We start by using (\ref{Ejjumps}) to see that
\[
\langle f,\mu^N_t\rangle_E = \langle f,\mu^N_0\rangle_E + A^{1,N}_t
+ A^{2,N}_t +
B^{1,N}_t + B^{2,N}_t,
\]
where
\begin{eqnarray*}
A^{1,N}_t &= &\frac{1}{N}\sum_{n=1}^N\int_{s=0}^t a^{1,N,n}_s \,ds,\\
A^{2,N}_t &= &\sum_{n=1}^N \int_{s=0}^t \jump^f_\NN(s)\,d(1-\dfi^\NN_s),
\\
B^{1,N}_t &= &\frac{1}{N}\sum_{n=1}^N \int_{s=0}^t \sigma_\NN\,\frac
{\partial f}{\partial\lambda}(\hat\mathsf{p}^\NN_s)\sqrt{\lambda
^\NN_s}\dfi
^\NN_s \,dW^n_s, \\
B^{2,N}_t &= &\eps_N\frac{1}{N}\sum_{n=1}^N \int_{s=0}^t \beta
^S_\NN
\lambda^\NN_s\,\frac{\partial f}{\partial\lambda}(\hat\mathsf
{p}^\NN_s)\dfi^\NN
_s \,dV_s,
\end{eqnarray*}
where, for simplicity, we have defined
\begin{eqnarray*} a^{1,N,n}_s &\Def&\frac12\biggl\{\bigl(\sigma^2_\NN
\lambda
^\NN_s + \eps_N^2(\beta^S_\NN)^2 (\lambda^\NN_s)^2
\bigr)\,\frac{\partial^2 f}{\partial\lambda^2}(\hat\mathsf{p}^\NN
_s)\\
&&\hspace*{12.5pt}{}+ \bigl(- \alpha_\NN(\lambda^\NN_s - \bar\lambda
_\NN
)- \eps_N \beta^S_\NN\lambda^\NN_sX_s\bigr)\,\frac{\partial
f}{\partial
\lambda}(\hat\mathsf{p}^\NN_s)\biggr\}\dfi^\NN_s.
\end{eqnarray*}
Thus, for any $0\le s\le t\le T$,
\begin{eqnarray*}
&&\BE[q^2(\langle f,\mu^N_t\rangle_E,\langle
f,\mu^N_s\rangle_E)
| \filt^N_s] \\
&&\qquad \le4\{\BE[q^2(A^{1,N}_t,A^{1,N}_s)| \filt
^N_s
] +\BE[q^2(A^{2,N}_t,A^{2,N}_s)| \filt^N_s]
\\
&&\hspace*{7pt}\qquad\quad{} +\BE[q^2(B^{1,N}_t,B^{1,N}_s)| \filt
^N_s] +\BE[q^2(B^{2,N}_t,B^{2,N}_s)| \filt
^N_s]\}
\\
&&\qquad \le 4\{\BE[|A^{1,N}_t-A^{1,N}_s|| \filt^N_s]
+\BE[|A^{2,N}_t-A^{2,N}_s|| \filt^N_s] \\
&&\hspace*{7pt}\qquad\quad{} +\BE[|B^{1,N}_t-B^{1,N}_s
|^2
| \filt^N_s] +\BE[|B^{2,N}_t-B^{2,N}_s
|^2|
\filt^N_s]\}.
\end{eqnarray*}

We now need to get some bounds.
Due to Assumption~\ref{ABounded}, for any $0\le s\le t< T$, we have that
\[
|\jump^f_\NN(t)|\le\frac1N\biggl\{\KK_{\mmbox{\ref{ABounded}}}\biggl\|\frac
{\partial f}{\partial\lambda}\biggr\|_C +\|f\|\biggr\}.
\]
This implies
\[
|A^{2,N}_t-A^{2,N}_s|\le\biggl\{\KK_{\mmbox{\ref{ABounded}}}\biggl\|\frac
{\partial
f}{\partial\lambda}\biggr\|_C +\|f\|\biggr\}|L^N_t-L^N_s|;
\]
thus, by Lemma~\ref{Ltight} we have that
\[
\BE[|A^{2,N}_t-A^{2,N}_s| |\filt^N_s]\le(t-s)^{1/4}\biggl\{\KK_{\mmbox{\ref
{ABounded}}}\biggl\|\frac{\partial f}{\partial\lambda}\biggr\|_C +\|
f\|
\biggr\}\BE[\Xi_N|\filt^N_s]
\]
for all $0\le s\le t\le T$.
To bound the increments of $A^{1,N}$, define
\[
\Xi^{(1)}_N \Def\frac12\Biggl\{1 + \frac{1}{N}\sum_{n=1}^N \int_{r=0}^t
(a^{1,N,n}_r)^2\,dr\Biggr\}.
\]
By Lemmata~\ref{Lmacrobound} and~\ref{LTechnicalBounds1} we have that
$\sup_{N\in\N}\BE[\Xi_N^{(1)}]<\infty$. By (\ref{Eintbnd}),
\[
|A^{1,N}_t-A^{1,N}_s|\le(t-s)^{1/4}\BE\bigl[\Xi^{(1)}_N|\filt^N_s\bigr].
\]

We next turn to the martingale terms. We have that
\begin{eqnarray*}
&&
\BE[|B^{1,N}_t - B^{1,N}_s|^2 | \filt
^N_s]\\
&&\qquad=\BE\Biggl[ \frac1N\sum_{n=1}^N \int_{r=s}^t \biggl(\sigma_\NN
\,\frac
{\partial f}{\partial\lambda}(\hat\mathsf{p}^\NN_r)\sqrt{\lambda
^\NN_r}\dfi
^\NN_r\biggr)^2 \,dr \Big| \filt^N_s\Biggr] \\
&&\qquad\le\BE\Biggl[ \frac1N\sum_{n=1}^N \int_{r=s}^t \biggl(\sigma_\NN
\,\frac
{\partial f}{\partial\lambda}(\hat\mathsf{p}^\NN_r)
\biggr)^2\lambda^\NN_r \,dr
\Big| \filt^N_s\Biggr] \\
&&\qquad\le(t-s)^{1/4}\BE\bigl[\Xi^{(2)}_N| \filt^N_s\bigr], \\
&&\BE[|B^{2,N}_t-B^{2,N}_s|^2 | \filt^N_s] \\
&&\qquad=\eps
_N^2 \BE
\Biggl[ \int_{r=s}^t \Biggl(\frac1N\sum_{n=1}^N \beta^S_\NN\lambda
^\NN
_r\,\frac{\partial f}{\partial\lambda}(\hat\mathsf{p}^\NN_s)\dfi
^\NN_r
\Biggr)^2 \,dr \Big| \filt^N_s\Biggr] \\
&&\qquad\le\eps_N^2\BE\Biggl[ \frac1N\sum_{n=1}^N \int_{r=s}^t
\biggl(\beta
^S_\NN\,\frac{\partial f}{\partial\lambda}(\hat\mathsf{p}^\NN
_r)\biggr)^2
(\lambda^\NN_r)^2 \,dr \Big| \filt^N_s\Biggr] \\
&&\qquad\le\eps_N^2(t-s)^{1/4}\BE\bigl[\Xi^{(2)}_N| \filt^N_s\bigr],
\end{eqnarray*}
where
\begin{eqnarray*} \Xi^{(2)}_N&\Def&\frac12 \Biggl\{1+ \frac1N\sum_{n=1}^N
\int
_{r=0}^T\biggl(\sigma_\NN\,\frac{\partial f}{\partial\lambda}(\hat
\mathsf{p}^\NN
_r)\biggr)^4(\lambda^\NN_r)^2 \,dr\Biggr\},\\
\Xi^{(3)}_N &\Def&\frac{1}{2} \Biggl\{1+ \frac1N\sum_{n=1}^N \int
_{r=0}^T\biggl(\beta^S_\NN\,\frac{\partial f}{\partial\lambda}(\hat
\mathsf{p}
^\NN_r)\biggr)^4(\lambda^\NN_r)^4 \,dr\Biggr\}.
\end{eqnarray*}
We have that
\[
\sup_{N\in\N}\BE\bigl[\Xi^{(2)}_N\bigr]<\infty\quad\mbox{and}\quad \sup
_{N\in\N}\BE\bigl[\Xi^{(3)}_N\bigr]<\infty.
\]
Collecting things together, we get that for any $0\le t\le T$, $0\le
u\le\delta$, and $0\le v\le\delta\wedge t$,
\begin{eqnarray*}
&&
\BE[q^2(\langle f,\mu^N_{t+u}\rangle
_E,\langle f,\mu^N_t\rangle
_E)| \filt^N_t]q^2(\langle f,\mu^N_t\rangle_E,\langle
f,\mu^N_{t-v}\rangle_E)
\\
&&\qquad \le \BE[q^2(\langle f,\mu^N_{t+u}\rangle_E,\langle
f,\mu^N_t\rangle_E)
| \filt^N_t] \\
&&\qquad \le4 \delta^{1/4}\BE\biggl[\biggl\{\Xi^{(1)}_N + \biggl\{\KK_{\mmbox{\ref
{ABounded}}}\biggl\|\frac{\partial f}{\partial\lambda}\biggr\|_C +\|
f\|
\biggr\}\Xi_N + \Xi^{(2)}_N + \eps_N^2 \Xi^{(3)}_N\biggr\}\Big| \filt
^N_t
\biggr].\qquad
\end{eqnarray*}
\upqed\end{pf}

We can now prove the desired relative compactness.
%
\begin{lemma}\label{LMuMeasureTight} The sequence $\{\mu^N\}_{N\in
\N
}$ is relatively compact in $D_{E}[0,\infty)$.\vadjust{\goodbreak}
\end{lemma}
\begin{pf}
Given Lemmas~\ref{LCompactContainment} and~\ref{LRegularity}, the
statement follows by Theorem~8.6 of Chapter 3 of~\cite{MR88a60130}.
\end{pf}

\section{Uniqueness}\label{SUniqueness}
We next verify that the solution of the resulting martingale problem is
\textit{unique}. We will use a duality argument (cf. Chapter 4.4 of~\cite{MR88a60130}).
In particular, here duality means that existence of a
solution to a dual problem ensures uniqueness to the original problem.
%
\begin{lemma}[(Uniqueness)]\label{Luniq} There is at most one solution
of the martingale problem for $\genA$ of (\ref{Elimgen}) with initial
condition $\pi\times\Lambda_\circ$.
\end{lemma}
\begin{pf} We will use the duality arguments of Chapter 4.4 of \cite
{MR88a60130}. Define $E^*\Def\bigcup_{M=1}^\infty C^\infty(\hat
P^M)$. Let's begin by defining a flow on $E^*$
as follows. Fix $f\in E^*$. Then $f\in C^\infty(\hat P^M)$ for some
$M\in\N$. Fix next $(\hat\mathsf{p}_1,\hat\mathsf{p}_2 ,\ldots,\hat
\mathsf{p}_M)\in\hat
\PP^M$
where $\hat\mathsf{p}_m=(\mathsf{p}_m,\lambda_m)$ and $\mathsf
{p}_m=(\alpha_m,\bar\lambda
_m,\sigma_m,\beta^C_m,\beta^S_m)$ for $m\in\{1,2,\ldots, M\}$. Define
\[
(T_t f)(\hat\mathsf{p}_1,\hat\mathsf{p}_2,\ldots,\hat\mathsf{p}_M)
\Def\BE\Biggl[f(\hat\mathsf{p}
^{*,1}_t,\hat\mathsf{p}^{*,2}_t,\ldots,\hat\mathsf{p}^{*,M}_t)\exp
\Biggl[-\sum
_{m=1}^M\int_{s=0}^t \lambda^{*,m}_s\,ds\Biggr]\Biggr],
\]
where $\hat\mathsf{p}^{*,m}_t = (\mathsf{p},\lambda^{*,m}_t)$ and
\[
\lambda^{*,m}_t = \lambda_m -\alpha_m \int_{s=0}^t (\lambda
^{*,m}_s-\bar\lambda_m)\,ds+ \sigma_m\int_{s=0}^t \sqrt{\lambda
^{*,m}_s}\,dW^m_s
\]
for all $m\in\{1,2,\ldots, M\}$. We also define
\[
(H_m f)(\hat\mathsf{p}_1,\hat\mathsf{p}_2, \ldots,
\hat\mathsf{p}_M,\hat\mathsf{p}_{M+1}) = M\beta
^C_{M+1}\lambda_{M+1}\,\frac{\partial f}{\partial\lambda_{m}}(\hat
\mathsf{p}
_1,\hat\mathsf{p}_2, \ldots, \hat\mathsf{p}_M)
\]
for\vspace*{2pt} $m\in\{1,2,\ldots,M\}$ and
$\hat\mathsf{p}_{M+1}=(\mathsf {p}_{M+1},\lambda _{M+1})\in\hat\PP$
where $\mathsf{p}_{M+1}=(\alpha_{M+1},\break\bar \lambda
_{M+1},\sigma_{M+1},\beta^C_{M+1},\beta^S_{M+1})$. Suppose\vspace*{-1pt} that $f\in
E^*$ and that in fact $f\in C^\infty(\hat P^M)$ for some $M\in\N$. Let
$\ee$ be an exponential(1) random variable. Set $F_t\Def T_t f$ for
\mbox{$t<\ee$}. Select $m\in\{1,2,\ldots,M\}$ according to a uniform
distribution on $\{1,2,\ldots, M\}$ and set $F_\ee\Def H_{m}(T_\ee f)$.
Restart the system.

Let's now connect $F$ to $\mu$. Fix $f\in E^*$ and $\mu\in E$. Then
$f\in C^\infty(\hat\PP^M)$ for some $M\in\N$, and we define
%
\begin{equation}\label{Ephidef} \phi(\mu,f) \Def\int_{(\hat
\mathsf{p}
_1,\hat\mathsf{p}_2,\ldots,\hat\mathsf{p}_M)\in\hat\PP^M}f(\hat
\mathsf{p}_1,\hat\mathsf{p}
_2,\ldots,\hat\mathsf{p}_M)\mu(d\hat\mathsf{p}_1)\mu(d\hat\mathsf
{p}_2)\cdots\mu(d\hat\mathsf{p}
_M).\hspace*{-28pt}
\end{equation}
If we fix $1=m_1<m_2<m_3<\cdots< m_{L+1}=M+1$ and $\{\tilde f_l\}
_{l=1}^L\subset C^\infty(\hat\PP)$ and assume that
\[
f(\hat\mathsf{p}_1,\hat\mathsf{p}_2,\ldots,\hat\mathsf{p}_M) =
\prod_{1\le l\le L}\biggl\{\prod
_{m_l\le m<m_{l+1}-1} \tilde f_l(\hat\mathsf{p}_m)\biggr\}\vadjust{\goodbreak}
\]
for all $(\hat\mathsf{p}_1,\hat\mathsf{p}_2,\ldots,\hat\mathsf
{p}_M)\in\hat\PP^M$, then
\[
\phi(\mu,f) = \prod_{l=1}^L \langle\tilde f_l,\mu\rangle
_E^{m_{l+1}-m_l-1}.
\]
By Stone--Weierstrass, we can thus approximate $\Phi$ in $\SSS$ by
linear combinations of functions of the form $\phi(\cdot,f)$ of (\ref
{Ephidef}) for some $f$'s in $E$.

To proceed, let's fix $f\in E$ and apply $\genA$ to the function $\mu
\mapsto\phi(\mu,f)$ given by (\ref{Ephidef}). It is fairly easy to
see that if $\{\bar\mu^*_t\}_{t\ge0}$ satisfies the martingale
problem for $\genA$, then
for each $f\in E$,
\[
\varphi(\bar\mu^*_t,f) = \int_{s=0}^t h_1(\bar\mu^*_s,f)\,ds +
\mart
^{(1)}_t,
\]
where $\mart^{(1)}$ is a martingale and where, if $f\in C^\infty(\hat
\PP^M)$,
\begin{eqnarray*}
h_1(\mu,f) & = & \sum_{m=1}^M\int_{\hat\mathsf{p}=(\hat\mathsf
{p}_1,\hat\mathsf{p}
_2,\ldots,\hat\mathsf{p}_M)\in\PP^M}\{(\genL_{1,m} f)(\hat\mathsf
{p})+\langle\QQ,\mu\rangle_E(\genL_{2,m} f)(\hat\mathsf
{p})\}\\
&&\hspace*{100pt}{}\times\mu(d\hat\mathsf{p}
_1)\mu(d\hat\mathsf{p}_2)\cdots\mu(d\hat\mathsf{p}_M),
\end{eqnarray*}
where $\genL_{1,m}$ and $\genL_{2,m}$ denote, respectively, the actions
of $\genL_1$ and $\genL_2$ defined by (\ref{EOperators1}) on the $m$th
coordinate of $f$.
On the other hand, we also have that for $\mu\in E$,
\[
\varphi(\mu,F_t) = \int_{s=0}^t h_2(\mu,F_s)\,ds + \mart^{(2)}_t,
\]
where $\mart^{(2)}$ is a martingale and
\begin{eqnarray*}
h_2(\mu,f) & = & \sum_{m=1}^M\int_{\hat\mathsf{p}=(\hat\mathsf
{p}_1,\hat\mathsf{p}
_2,\ldots,\hat\mathsf{p}_M)\in\PP^M}(\genL_{1,m} f)(\hat\mathsf
{p})\mu(d\hat\mathsf{p}
_2)\cdots\mu(d\hat\mathsf{p}_M) \\
&&{} + \frac{1}{M}\sum_{m=1}^M\{\varphi(\mu,H_m f)-\varphi(\mu,f)\}.
\end{eqnarray*}
Note that
\begin{eqnarray*}
&&\frac1M\sum_{m=1}^M\varphi(\mu,H_m f)\\
&&\qquad= \sum_{m=1}^M\int_{\hat\mathsf{p}=(\hat\mathsf{p}_1,\hat
\mathsf{p}_2,\ldots,
\hat\mathsf{p}_M,\hat\mathsf{p}_{M+1})\in\PP^{M+1}}\beta
^C_{M+1}\lambda_{M+1}\,\frac
{\partial f}{\partial\lambda_{m}}(\hat\mathsf{p})\mu(d\hat\mathsf
{p}_2)\cdots\mu
(d\hat\mathsf{p}_{M+1})\\
&&\qquad= \sum_{m=1}^M\int_{\hat\mathsf{p}=(\hat\mathsf{p}_1,\hat
\mathsf{p}_2,\ldots,
\hat\mathsf{p}_M)\in\PP^M}\langle\QQ,\mu\rangle_E (\genL_{2,m}
f)(\hat\mathsf{p})\mu
(d\hat\mathsf{p}_2)\cdots\mu(d\hat\mathsf{p}_M).
\end{eqnarray*}

Collecting things together, we have that
\[
h_1(\mu,f) = h_2(\mu,f) +\varphi(\mu,f)
\]
and this implies uniqueness.\vspace*{-2pt}
\end{pf}

\section{Proof of main theorem}\label{SMainProof}
We now have our first convergence result. Let $\mathbb{Q}_N$ be the
$\BP
$-law of $\mu^N$, that is,
\[
\mathbb{Q}_N(A) \Def\BP\{\mu^N\in A\}
\]
for all $A\in\Borel(D_E[0,\infty))$. Thus, $\mathbb{Q}_N\in
\Pspace(D_E[0,\infty))$ for all $N\in\N$. For $\omega\in
D_E[0,\infty)$, define $X_t(\omega)\Def\omega(t)$ for
all $t\ge0$.\vspace*{-2pt}
%
\begin{proposition}\label{Pconv} We have that $\mathbb{Q}_N$ converges
[in the topology of\break $\Pspace(D_E[0$, $\infty))$] to the
solution $\mathbb{Q}$ of the martingale problem generated by $\genA$ of
(\ref{Elimgen}) and such that $\mathbb{Q} X_0^{-1} = \delta_{\pi
\times
\Lambda_\circ}$.
In other words, $\mathbb{Q}\{X_0=\pi\times\Lambda_\circ\}=1$ and
for all $\Phi\in\SSS$ and $0\le r_1\le r_2\le\cdots\le r_J=s<t<T$ and
$\{\psi_j\}_{j=1}^J\subset B(E)$, we have that
\[
\lim_{N\to\infty} \BE^{\mathbb{Q}}\Biggl[\biggl\{\Phi(X_t)-\Phi
(X_s)-\int
_{r=s}^t (\genA\Phi)(X_r)\,dr\biggr\}\prod_{j=1}^J \psi_j(X_{r_j})\Biggr]=0,
\]
where $\BE^{\mathbb{Q}}$ is the expectation operator defined by
$\mathbb{Q}$.\vspace*{-2pt}
\end{proposition}
\begin{pf}
The result follows from Lemmata~\ref{Lwconv},~\ref{LMuMeasureTight}
and~\ref{Luniq}. Of course, we also have that for any $\Phi\in\SSS$,
\[
\BE^{\mathbb{Q}}[\Phi(X_0)] = \lim_{N\to\infty}\BE^{\mathbb
{Q}}[\Phi
(\mu^N_0)] = \Phi(\pi\times\Lambda_\circ),
\]
which implies the claimed initial condition.\vspace*{-2pt}
\end{pf}

We next want to identify $\mathbb{Q}$.\vspace*{-2pt}
%
\begin{lemma}\label{LQchar} We have that $\mathbb{Q}= \delta_{\bar
\mu}$, where $\bar\mu$ is given by (\ref{Emudef}).\vspace*{-2pt}
\end{lemma}
\begin{pf} Recall (\ref{EEffectiveEquation1}) and the operators
$\genL
_1,\genL_2$ from (\ref{EOperators1}) and the definition of $Q$ in
(\ref{EbQDef}). For any $f\in C^\infty(\hat\PP)$,
\[
\langle f,\bar\mu_t\rangle_E = \int_{\hat\mathsf{p}= (\mathsf
{p},\lambda)\in\hat\PP} \BE
\biggl[f(\mathsf{p},\lambda_t^*(\hat\mathsf{p}))\exp\biggl[-\int
_{s=0}^t \lambda
_s^*(\hat\mathsf{p})\,ds\biggr]\biggr] \pi(d\mathsf{p})\Lambda
_\circ(d\lambda).
\]
Thus,
\begin{eqnarray*} \frac{d}{dt}\langle f,\bar\mu_t\rangle_E &= &\int
_{\hat\mathsf{p}= (\mathsf{p}
,\lambda)\in\hat\PP} \BE\biggl[(\genL_1 f)(\mathsf{p},\lambda
^*_t(\hat\mathsf{p}
))\exp\biggl[-\int_{s=0}^t \lambda_s^*(\hat\mathsf{p})\,ds
\biggr]\biggr] \pi(d\mathsf{p}
)\Lambda_\circ(d\lambda)\\
&&{}+ \int_{\hat\mathsf{p}= (\mathsf{p},\lambda)\in\hat\PP} \BE
\biggl[(\genL_2 f)(\mathsf{p}
,\lambda^*_t(\hat\mathsf{p}))Q(t)\exp\biggl[-\int_{s=0}^t \lambda
_s^*(\hat\mathsf{p}
)\,ds\biggr]\biggr] \vadjust{\goodbreak}\\
&&\hspace*{58pt}{}\times\pi(d\mathsf{p})\Lambda_\circ(d\lambda)\\
&= &\langle\genL_1f,\bar\mu_t\rangle_E+ Q(t) \langle\genL_2f,\bar
\mu_t\rangle_E.
\end{eqnarray*}
To proceed, define
\[
G(t)\Def\mathop{\int_{\hat\mathsf{p}=(\mathsf{p},\lambda)\in
\hat\PP}}_{
\mathsf{p}=(\alpha,\bar\lambda,\sigma,\beta^C,\beta^S)} \beta^C
\BE\biggl[\exp
\biggl[-\int_{s=0}^t \lambda_s^*(\hat\mathsf{p})\,ds\biggr]\biggr]
\pi(d\mathsf{p}
)\Lambda_\circ(d\lambda).
\]
On the one hand, we have that
\begin{eqnarray*} \dot G(t) &=& -\mathop{\int_{\hat\mathsf
{p}=(\mathsf{p},\lambda)\in
\hat\PP}}_{
\mathsf{p}=(\alpha,\bar\lambda,\sigma,\beta^C,\beta^S)} \beta^C
\BE
\biggl[\lambda_t^*(\hat\mathsf{p})\exp\biggl[-\int_{s=0}^t \lambda
_s^*(\hat\mathsf{p}
)\,ds\biggr]\biggr] \pi(d\mathsf{p})\Lambda_\circ(d\lambda)\\
&=& -\mathop{\int_{\hat\mathsf{p}=(\mathsf{p},\lambda)\in\hat\PP}}_{
\mathsf{p}=(\alpha,\bar\lambda,\sigma,\beta^C,\beta^S)} \beta^C
\lambda\bar
\mu_t(d\hat\mathsf{p}) = -\langle\QQ,\bar\mu_t\rangle_E.
\end{eqnarray*}
We want to show that
%
\begin{equation}\label{Egg} \dot G(t) = -Q(t).
\end{equation}
Indeed, fix $\hat\mathsf{p}=(\mathsf{p},\lambda)\in\hat\PP$
where $\mathsf{p}=(\alpha
,\bar\lambda,\sigma,\beta^C,\beta^S)$. Define
\[
M_s\Def\exp\biggl[-b^\mathsf{p}(t-s)\lambda_s^*(\hat\mathsf{p}) -
\int_{r=s}^t b^\mathsf{p}
(t-r)\{ Q(r) + \alpha\bar\lambda\} \,dr - \int_{r=0}^s \lambda
_r^*(\hat\mathsf{p})\,dr\biggr]
\]
for $0\le s\le t$. Using the calculations of~\cite{DuffiePanSingleton},
\begin{eqnarray*} dM_s &=& d\mart_s + \bigl\{\dot b^\mathsf{p}(t-s)\lambda
_s^*(\hat
\mathsf{p}) - b^\mathsf{p}(t-s)\bigl\{-\alpha\bigl(\lambda_s^*(\hat\mathsf
{p})-\bar\lambda\bigr) +
Q(s)\bigr\}\\
&&\hspace*{38.7pt}{} + \tfrac12\sigma^2\bigl(b^\mathsf{p}(t-s)\bigr)^2\lambda
_s^*(\hat\mathsf{p}) +
b^\mathsf{p}(t-s)\bigl(Q(s) + \alpha\bar\lambda\bigr)-\lambda
_s^*(\hat\mathsf{p}
)\bigr\} M_s \,ds \\
&=& d\mart_s,
\end{eqnarray*}
where $\mart$ is a martingale [we use here the ODE (\ref
{EDuffiePanSingleton})]. Noting that
\begin{eqnarray*} M_0 &= &\exp\biggl[-b^\mathsf{p}(t)\lambda- \int
_{r=0}^t b^\mathsf{p}
(t-r)\{ Q(r) + \alpha\bar\lambda\} \,dr \biggr],\\
M_t &= &\exp\biggl[- \int_{r=0}^t \lambda_r^*(\hat\mathsf
{p})\,dr\biggr],
\end{eqnarray*}
we have that
\begin{eqnarray*}
G(t) & = & \mathop{\int_{\hat\mathsf{p}=(\mathsf{p},\lambda)\in
\hat\PP}}_{
\mathsf{p}=(\alpha,\bar\lambda,\sigma,\beta^C,\beta^S)} \beta^C
\exp
\biggl[-b^\mathsf{p}(t)\lambda\\
&&\hspace*{99pt}{} - \int_{r=0}^t b^\mathsf{p}(t-r) \{ Q(r) + \alpha\bar\lambda\}
\,dr
\biggr]\\
&&\hspace*{61.5pt}{}\times\pi(d\mathsf{p})\Lambda_\circ(d\lambda).
\end{eqnarray*}
Differentiating this, we get that
\begin{eqnarray*} \dot G(t) &=& -\mathop{\int_{\hat\mathsf
{p}=(\mathsf{p},\lambda)\in
\hat\PP}}_{
\mathsf{p}=(\alpha,\bar\lambda,\sigma,\beta^C,\beta^S)} \beta^C
\biggl[ \dot
b^\mathsf{p}(t)\lambda+ \int_{r=0}^t\dot b^\mathsf{p}(t-r)\{ Q(r)+
\alpha\bar
\lambda\} \,dr\biggr]\\[-2pt]
&&\hspace*{71pt}{} \times\exp\biggl[-b^\mathsf{p}(t)\lambda- \int_{r=0}^t
b^\mathsf{p}(t-r)\{
Q(r) + \alpha\bar\lambda\} \,dr \biggr]\\[-2pt]
&&\hspace*{71pt}{}\times\pi(d\mathsf{p})\Lambda
_\circ(d\lambda
) \\[-2pt]
&=& -Q(t),
\end{eqnarray*}
where we have used the defining equation (\ref{EbQDef}) for $Q$.
Thus, (\ref{Egg}) holds, so we have that
\[
\frac{d}{dt}\langle f,\bar\mu_t\rangle_E = \langle\genL_1f,\bar
\mu_t\rangle_E+ \langle
\QQ,\bar\mu_t\rangle_E \langle\genL_2f,\bar\mu_t\rangle_E.
\]
Thus,
\[
\Phi(\bar\mu_t) = \Phi(\bar\mu_0)+\int_{s=0}^t (\genA\Phi
)(\bar\mu
_s)\,ds,
\]
and, hence, $\delta_{\bar\mu}$ satisfies the martingale problem
generated by $\genA$. Of course, we also have that $\bar\mu_0 = \pi
\times\Lambda_\circ$. By uniqueness, the claim follows.
\end{pf}

We now can finish the proof of our main result.
\begin{pf*}{Proof of Theorem~\ref{TMainLLN}}
Since weak convergence to a constant implies convergence in
probability, we have (\ref{Emulimit}). Using\vspace*{1pt} the
fact that the map $\varphi\dvtx\hat\PP\mapsto1$ is in $C(\hat \PP)$,
$L^N$ is a continuous transformation of $\mu^N$ into $D_\R[0,\infty)$.
From (\ref{EFrep}) we have that
\[
\lim_{N\to\infty}\BP\{ d_\R(L^N,F)\ge\delta\}= 0
\]
for each $\delta>0$. To finish the proof, we need to replace the
Skorohod norm $d_\R$ by the supremum norm.\vspace*{1pt} From (\ref{EFDer}) we have
that $\KK_T\Def\sup_{0\le t\le T}\dot F(t)$ is finite for each $T>0$.
To get the claimed convergence, we adopt the notation of Chapter 3.5 of
\cite{MR88a60130}.
For any nondecreasing and differentiable map $g$ of $[0,T]$ into
itself and any $t\in[0,T]$, we have that
\begin{eqnarray*} |L^N_t-F(t)| &\le&|L^N_t-F(g(t))| + |F(g(t))-F(t)|\\
&\le&\sup_{0\le t\leq T}|L^N_t-F(g(t))| + K_T|g(t)-t|\\
&\le&\sup_{0\le t\leq T}|L^N_t-F(g(t))| + K_T T \sup_{0\le t\leq
T}|\dot g(t)-1|\\
&\le&\sup_{0\le t\leq T}|L^N_t-F(g(t))| \\
&&{} + K_T T \max\Bigl\{\Bigl|\exp\Bigl[\sup_{0\leq t\leq T}|{\log
\dot g(t)}|\Bigr]-1\Bigr|, \\
&&\hspace*{62.8pt} \Bigl|\exp\Bigl[-\sup_{0\le t\leq T}|{\log
\dot g(t)}|\Bigr]-1\Bigr|\Bigr\}.
\end{eqnarray*}
Varying $g$, we get that
\begin{eqnarray*}
&&
\sup_{0\le t\leq T} |L^N_t-F(t)| \\
&&\qquad \le d_{\R}(L^N,F)
+ \KK_T T \max
\{|{\exp}[d_{\R}(L^N,F)]-1|,\\
&&\hspace*{48.6pt}\hspace*{94pt}|{\exp}[-d_{\R}(L^N,F)]-1|\}.
\end{eqnarray*}
The claim now follows; note that $F$ and $L^N$ both take values in
$[0,1] $.
\end{pf*}

\section{Conclusion and extensions}\label{Sconclusion}
We have developed a point process model of correlated default timing in
a portfolio of firms, and have analyzed typical default profiles in the
limit as the size of the pool grows. Our empirically motivated model
captures two important sources of default clustering, namely, the
exposure of firms to a systematic risk process, and contagion. We have
proved a law of large numbers for the default rate in the pool.

There are several potential extensions of our work. For example, the
default intensity dynamics (\ref{Emain}) can be generalized to include
a dependence on the systematic risk process of the magnitude of the
jump at a default. Then, the impact of a default on the surviving firms
depends on the state of the systematic risk: intuitively, if the
economy is weak, firms are fragile and more susceptible to contagion.
This generalization of the intensity dynamics is empirically plausible,
and can be treated with arguments similar to the ones we currently use.

\section{\texorpdfstring{Proofs of Lemmas \protect\ref{Lexistence}, \protect\ref{Luniqueness}, \protect\ref{Lmacrobound} and \protect\ref{LbQDef}}
{Proofs of Lemmas 3.1, 3.2, 3.4 and 4.1}}\label{SAppendix}

In this section we prove Lemmas~\ref{Lexistence},~\ref{Luniqueness},
\ref{Lmacrobound} and~\ref{LbQDef}. For presentation purposes, we first
collect in Lemma~\ref{LTechnicalBounds1} some a-priori bounds that will
be useful in the proof of these lemmas. Then, in Section~\ref{SSBounds}
we proceed with the proof of Lemmas~\ref{Lexistence},~\ref{Luniqueness}
and~\ref{Lmacrobound}. We mention here that the square-root singularity
unavoidably complicates the analysis. The theory behind CIR-like
processes is a bit delicate due to the square root singularity in the
diffusion, so we need to develop some new modifications to existing
results (cf.~\cite{MR92h60127,MR1011252,MR92d60053}). Last, in
Section~\ref{SSBounds2} we prove Lemma~\ref{LbQDef}.

\subsection{Effect of systematic risk}
Our first step is to get some usable bounds on the systematic risk $X$.
We need these bounds since, as we mentioned in Section~\ref
{SWellposednessProperties}, the $\lambda_t \,dX_t$ term contains the
term $\lambda_t X_t \,dt$, implying that the dynamics of the $\R
^2$-valued process $(\lambda,X)$ contain a superlinear drift. Note that
the systematic risk process $X$ of course has an explicit form:
\[
X_t = e^{-\gamma t}x_\circ+ \int_{s=0}^t e^{-\gamma(t-s)}\,dV_s,\qquad
t>0.
\]
Fix $\mathsf{p}=(\alpha,\bar\lambda,\sigma,\beta^C,\beta^S)\in
\PP$, $\lambda
_\circ$ in $\R_+$, and $\xi$ as required in the beginning of Section
\ref{SWellposednessProperties}. Define
\begin{eqnarray*} \Gamma_t &\Def&\alpha t +\beta^S \gamma\int_{s=0}^t
X_s\,ds, \\
Z_t&\Def&\lambda_\circ+\alpha\bar\lambda\int_{s=0}^t e^{\Gamma
_s}\,ds+ \beta^C\int_{s=0}^t e^{\Gamma_s}\,d\xi_s\\
&= &\lambda_\circ+\alpha\bar\lambda\int_{s=0}^t e^{\Gamma_s}\,ds+
\beta
^C\biggl\{ e^{\Gamma_t} \xi_t - \int_{s=0}^t e^{\Gamma_s}\xi_s( \alpha+
\beta^S \gamma X_s) \,ds\biggr\}\\
&= &\lambda_\circ+\int_{s=0}^t e^{\Gamma_s}\{\alpha\bar\lambda
-\beta^C\xi_s(\alpha+\beta^S \gamma X_s)\} \,ds+ \beta
^C\xi_t
e^{\Gamma_t}
\end{eqnarray*}
for all $t\ge0$. The alternate representations of $Z$ will allow us
bounds which are independent of $\xi$.

Our first result is a bounds on $X$, $\Gamma$ and $Z$ which explicitly
depend on various coefficients. The importance of the bound on the
moments of $Z_t$ is that they do not depend on $\xi$.
%
\begin{lemma}\label{LTechnicalBounds1} For each $p\ge1$ and $t\ge0$,
\begin{eqnarray*} \BE[X_t^{2p}]^{1/(2p)} &\le&|
x_\circ
| + \frac{1}{2\sqrt{\gamma}}\biggl( \frac{(2p)!}{p!}
\biggr)^{1/(2p)},\\
\BE[\exp[ p\Gamma_t]]&\le&\exp\bigl[ |p|
\{
\alpha t + |\beta^C x_\circ|\}+ \tfrac12(p\beta^C)^2t\bigr], \\
\BE[Z_t^{2p}]^{1/(2p)}&\le&\lambda_\circ+ |\beta^C|\BE
[e^{-2p\Gamma_t} ]^{1/(2p)} \\
&&{} +t^{1-1/(2p)}\biggl(\int_{s=0}^t \BE[e^{-4p\Gamma
_s}
]\,ds\biggr)^{1/(4p)}\\
&&\hspace*{10pt}{} \times\biggl\{\alpha\bar\lambda t^{1/(4p)} + \alpha|\beta
^C| t^{1/(4p)} \\
&&\hspace*{27pt}{} + |\beta^C\beta^S\gamma|\biggl(\int_{s=0}^t
\BE
[X_s^{4p}] \,ds\biggr)^{1/(4p)}\biggr\}.
\end{eqnarray*}
\end{lemma}
\begin{pf} We first bound $X$. For every $p\ge1$ and $t\ge0$
\begin{eqnarray*} \BE[X_t^{2p}]^{1/(2p)} &\le&|
x_\circ
e^{-\gamma t}| + \biggl\{\BE\biggl[\biggl|\int_{s=0}^t e^{-\gamma
(t-s)}\,dV_s\biggr|^{2p}\biggr]\biggr\}^{1/(2p)}\\
&=&| x_\circ e^{-\gamma t}| + \sqrt{\int_{s=0}^t
e^{-2\gamma
(t-s)}\,ds} \biggl( \frac{(2p)!}{2^p p!}\biggr)^{1/(2p)}\\
&\le&| x_\circ| + \frac{1}{\sqrt{2\gamma}}\biggl(
\frac
{(2p)!}{2^p p!}\biggr)^{1/(2p)}\\
&=& | x_\circ| + \frac{1}{2\sqrt{\gamma}}\biggl( \frac
{(2p)!}{p!}\biggr)^{1/(2p)}.
\end{eqnarray*}

Next note that
\begin{eqnarray*} \Gamma_t &= &\alpha t - \beta^S\gamma\int_{s=0}^t
x_\circ
e^{-\gamma s}\,ds\\
&&{} - \beta^S\gamma\int_{s=0}^t \biggl\{\int_{r=0}^s
e^{-\gamma
(s-r)} \,dV_r\biggr\} \,ds \\
&= &\alpha t - \beta^S x_\circ\{1-e^{-\gamma t}\}- \beta^S\gamma\int
_{r=0}^t \biggl\{\int_{s=r}^t e^{-\gamma(s-r)}\,ds\biggr\} \,dV_r \\
&= &\alpha t - \beta^S x_\circ\{1-e^{-\gamma t}\}- \beta^S\int
_{r=0}^t \bigl\{1-e^{-\gamma(t-r)}\bigr\} \,dV_r.
\end{eqnarray*}
Thus, for any $p\in\R$
\begin{eqnarray*} \BE[\exp[p\Gamma_t]] &=&
\exp\biggl[
p \{\alpha t + \beta^C x_\circ(1-e^{-\gamma t})\} \\
&&\hspace*{19.6pt}{} + \frac{(p\beta^C)^2}{2}\int_{r=0}^t \bigl\{1-e^{-\gamma
(t-r)}\bigr\}^2 \,dr \biggr]\\
&\le&\exp\biggl[ |p| \{\alpha t + |\beta^C x_\circ|\}+ \frac
12(p\beta
^C)^2t\biggr].
\end{eqnarray*}

We can finally bound $Z$. We have that
\begin{eqnarray*}
\BE[Z_t^{2p}]^{1/(2p)} & \le&\lambda_\circ+ \BE
\biggl[
\biggl(\int_{s=0}^t e^{-\Gamma_s}\{\alpha\bar\lambda-\beta^C\xi_s
(\alpha+\beta^S\gamma X_s)\} \,ds\biggr)^{2p}\biggr]^{1/(2p)}
\\
&&{} + |\beta^C|\BE[e^{-2p\Gamma_t} ]^{1/(2p)}.
\end{eqnarray*}
We also have that
\begin{eqnarray*}
&&\BE\biggl[\biggl(\int_{s=0}^t e^{\Gamma_s}\{
\alpha\bar
\lambda-\beta^C\xi_s(\alpha+\beta^S\gamma X_s)\}
\,ds
\biggr)^{2p}\biggr]^{1/(2p)}\\
&&\qquad \le \BE\biggl[\biggl(\int_{s=0}^t e^{2\Gamma_s}\,ds\biggr)^p
\biggl(\int_{s=0}^t \{\alpha\bar\lambda-\beta^C\xi_s(\alpha
+\beta
^S\gamma X_s)\}^2 \,ds\biggr)^p\biggr]^{1/(2p)}\\
&&\qquad \le \BE\biggl[\biggl(\int_{s=0}^t e^{2\Gamma_s}\,ds
\biggr)^{2p}
\biggr]^{1/(4p)} \\
&&\qquad\quad{} \times\BE\biggl[\biggl(\int_{s=0}^t \{\alpha\bar
\lambda-\beta^C\xi_s(\alpha+\beta^S \gamma X_s)\}
\,ds
\biggr)^{2p}\biggr]^{1/(4p)}\\
&&\qquad \le t^{1-1/(2p)}\BE\biggl[\int_{s=0}^t e^{-4p\Gamma
_s}\,ds
\biggr]^{1/(4p)} \\
&&\qquad\quad{} \times\BE\biggl[\int_{s=0}^t \{\alpha\bar\lambda
-\beta^C\xi_s(\alpha+\beta^S \gamma X_s)\}^{2p}
\,ds
\biggr]^{1/(4p)}\\
&&\qquad \le t^{1-1/(2p)}\biggl(\int_{s=0}^t \BE[e^{-4p\Gamma
_s}
]\,ds\biggr)^{1/(4p)}\\
&&\qquad\quad{}\times\biggl\{\alpha\bar\lambda t^{1/(4p)} + \alpha|\beta^C|
t^{1/(4p)}
+ |\beta^C\beta^S\gamma|\BE\biggl[\int_{s=0}^t
X_s^{4p} \,ds\biggr]^{1/(4p)}\biggr\}.
\end{eqnarray*}
Combine things together to get the bound on $Z$.
\end{pf}

\subsection{\texorpdfstring{Proofs of Lemmas \protect\ref{Lexistence}, \protect\ref{Luniqueness} and \protect\ref{Lmacrobound}}
{Proofs of Lemmas 3.1, 3.2 and 3.4}}\label{SSBounds}

Let's next understand the regularity of various CIR-like processes
which we use. Before proceeding with the proofs, we define a function
$\psi_\eta(x)$ that will be essential for the proofs. It is introduced
in order to deal with the square-root singularity. In particular, let
\[
\psi_\eta(x) \Def\frac{2}{\ln\eta^{-1}}\int_{y=0}^{|x|}\biggl\{\int
_{z=0}^y \frac{1}{z}\chi_{[\eta,\eta^{1/2}]}(z) \,dz\biggr\} \,dy
\quad\mbox
{and}\quad g_\eta(x) \Def|x|-\psi_\eta(x)
\]
for all $x\in\R$.

Let us then study some important properties of $\psi_\eta(x)$ that will
be repeatedly used in the proofs. First, we note that $\psi_\eta$ is
even, so $g_\eta$ is also even. Taking derivatives, we have that
\[
\dot\psi_\eta(x) = \frac{2}{\ln\eta^{-1}} \int_{z=0}^x \frac
{1}{z}\chi
_{[\eta,\eta^{1/2}]}(z) \,dz \quad\mbox{and}\quad \ddot\psi_\eta
(x) =
\frac{2}{\ln\eta^{-1}} \frac{1}{x}\chi_{[\eta,\eta^{1/2}]}(x)
\]
for all $x>0$. Since $\ddot g_\eta=-\ddot\psi_\eta\le0$, $\dot
g_\eta
$ is nonincreasing. For $x>\sqrt{\eta}$,
\[
\dot g_\eta(x) = 1-2\frac{\ln\eta^{1/2}-\ln\eta}{\ln
({1}/{\eta})}=0,
\]
so in fact $\dot g_\eta$ is nonnegative on $(0,\infty)$ and it
vanishes on $[\sqrt{\eta},\infty)$. Thus, $g_\eta$ is nondecreasing and
reaches its maximum at $\sqrt{\eta}$.
Since $g_\eta(0)=0$, we in fact have that
\[
0\le g_\eta(x)\le g_\eta\bigl(\sqrt{\eta}\bigr)
\]
for all $x\ge0$. Since $\dot g_\eta$ is nonincreasing on $(0,\infty)$
and $\dot g_\eta(x)=1$ for $x\in(0,\eta)$, we have that $\dot g_\eta
(x)\le1$
for all $x\in(0,\sqrt{\eta})$, so $g_\eta(\sqrt{\eta}) \le\sqrt
{\eta
}$. Since $g_\eta$ is even, we in fact must have that $|g_\eta(x)|\le
\sqrt{\eta}$ for all $x\in\R$.
Hence,
\[
|x|\le\psi_\eta(x)+\sqrt{\eta}
\]
for all $x\in\R$.
We finally note that
\[
|\ddot\psi_\eta(x)|\le\frac{2}{\ln\eta^{-1}}\frac
{1}{|x|}\chi_{[\eta,\infty)}(|x|)\le\frac{2}{\ln\eta^{-1}}\min\biggl\{
\frac{1}{|x|},\frac{1}{\eta}\biggr\}
\]
for all $x\in\R$.

Now we have all the necessary tools to proceed with the proof of the lemmas.
\begin{pf*}{Proof of Lemma~\ref{Lexistence}} For each $N\in\N$, define
\[
\vrho_N(t) \Def\frac{\lfloor tN\rfloor}{N}
\]
for all $t\in[0,T]$. For each $N\in\N$, define
\[
Y^N_t = \sigma\int_{s=0}^t e^{\Gamma_s/2}\sqrt{\bigl(Y^N_{\vrho
_N(s)}+Z_s\bigr)\vee0}\,dW^*_s + \beta\int_{s=0}^t\bigl(
\bigl(Y^N_{\vrho^N(s)} +Z_s\bigr)\vee0\bigr)\,dV_s.
\]
We will show that $(Z_t + Y^N_t)e^{\Gamma_t}$ converges to a solution
of (\ref{ElambdaSDE}) (as $N\nearrow\infty$).

As a first step, let's bound some moments. Fix $p>1$. For $0\le
s\le\break
t\le T$,~\cite{MR92h60127}, Exercise 3.25, gives us that
\begin{eqnarray*} &&\BE\biggl[\biggl|\int_{r=s}^t\bigl(
\bigl(Y^N_{\vrho
^N(r)}+Z_r\bigr) \vee0\bigr)\,dV_r\biggr|^{2p}\biggr] \\
&&\qquad \le \bigl(p(2p-1)\bigr)^p (t-s)^{p-1}\int_{r=s}^t \BE\bigl[
\bigl(
\bigl(Y^N_{\vrho^N(r)}+Z_r\bigr) \vee0\bigr)^{2p}\,dr\bigr]\\
&&\qquad \le \bigl(p(2p-1)\bigr)^p (t-s)^{p-1}\int_{r=s}^t \BE\bigl[
\bigl|Y^N_{\vrho^N(r)}+Z_r\bigr|^{2p}\bigr]\,dr \\
&&\qquad \le2^{2p-1}\bigl(p(2p-1)\bigr)^p (t-s)^{p-1}\\
&&\qquad\quad{}\times\biggl\{\int_{r=s}^t \BE
\bigl[\bigl|Y^N_{\vrho^N(r)}\bigr|^{2p}\bigr]\,dr
+ \int_{r=s}^t \BE[|Z_r
|^{2p}]\,dr\biggr\}.
\end{eqnarray*}
Similarly,
\begin{eqnarray*} &&\BE\biggl[\biggl|\int_{r=s}^t e^{\Gamma_r/2}\sqrt
{\bigl(Y^N_{\vrho_N(r)}+Z_r\bigr)\vee0}\,dW^*_r\biggr|^{2p}\biggr] \\
&&\qquad \le \bigl(p(2p-1)\bigr)^p (t-s)^{p-1}\int_{r=s}^t \BE
\bigl[e^{p\Gamma
_r}\bigl|\bigl(Y^N_{\vrho_N(r)}+Z_r\bigr)\vee0\bigr|^p
\bigr]\,dr\\
&&\qquad \le \frac12\bigl(p(2p-1)\bigr)^p (t-s)^{p-1}\biggl\{\int_{r=s}^t \BE
[e^{2p\Gamma_r}] + \BE\bigl[\bigl|Y^N_{\vrho
_N(r)}+Z_r
\bigr|^{2p}\bigr]\,dr\biggr\}\\
&&\qquad \le \frac12\bigl(p(2p-1)\bigr)^p (t-s)^{p-1}\\
&&\qquad\quad{}\times\biggl\{\int_{r=s}^t \BE
[e^{2p\Gamma_r}] \,dr+ 2^{2p-1}\int_{r=s}^t \BE\bigl[
\bigl|Y^N_{\vrho
_N(r)}\bigr|^{2p}\bigr]\,dr\\
&&\hspace*{142pt}{} + 2^{2p-1}\int_{r=s}^t \BE[
|Z_r|^{2p}]\,dr\biggr\}.
\end{eqnarray*}
We can bound the effect of $Z$ by Lemma~\ref{LTechnicalBounds1}.
Collecting things together, and using the fact that $\vrho_N(t)\le t$,
we have that there is a $\KK_A>0$ such that
\begin{eqnarray*}
\BE\bigl[\bigl|Y^N_{\vrho_N(t)}\bigr|^{2p}\bigr] & \le&\KK_A +
\KK_A
\int_{s=0}^{\vrho_N(t)} \BE\bigl[\bigl|Y^N_{\vrho_N(s)}
\bigr|^{2p}\bigr]\,dr \\
& \le&\KK_A + \KK_A \int_{s=0}^t \BE\bigl[\bigl|Y^N_{\vrho
_N(s)}
\bigr|^{2p}\bigr]\,dr
\end{eqnarray*}
for all $N\in\N$ and $t\in[0,T]$, which in turn implies that
%
\begin{equation}\label{Eaa} \sup_{0\le t\le T}\BE\bigl[
\bigl|Y^N_{\vrho
_N(t)}\bigr|^{2p}\bigr] \le\KK_A e^{\KK_A T}
\end{equation}
for $0\le t\le T$.
This in turn implies that there is a $\KK_B>0$ such that
%
\begin{equation}\label{Ebb} \BE\bigl[\bigl|Y^N_t-Y^N_{\vrho
^N(t)}
\bigr|^{2p}\bigr] \le\KK_B |t-\vrho_N(t)|^p\le\KK_B\frac{1}{N^p}
\end{equation}
for all $0\le t\le T$.

We next want to show that $Y^N$ converges in $L^1$. Fix $N$ and $N'$ in
$\N$ and define
\[
\nu^{N,N'}_t \Def Y^N_t-Y^{N'}_t.
\]
Fix also $\eta>0$. We have that
\[
|\nu^{N,N'}_t| \le\psi_\eta(\nu^{N,N'}_t) + \sqrt{\eta} = \sigma^2
A^{1,N,N'}_t+\beta^2 A^{2,N,N'}_t + \mart_t + \sqrt{\eta},
\]
where $\mart$ is a martingale and
\begin{eqnarray*} A^{1,N,N'}_t &=& \frac12 \int_{s=0}^t \ddot\psi
_\eta(\nu
^{N,N'}_s)e^{\Gamma_s}\bigl\{\sqrt{\bigl(Y^{N'}_{\vrho_N(s)}+Z_s
\bigr)\vee0} 
-\sqrt{\bigl(Y^N_{\vrho_{N'}(s)}+Z_s\bigr)\vee
0}\bigr\}^2
\,ds \\
&\le&\frac12 \int_{s=0}^t \ddot\psi_\eta(\nu^{N,N'}_s)e^{\Gamma
_s}
\bigl|Y^N_{\vrho_N(s)}-Y^{N'}_{\vrho_{N'}(s)}\bigr| \,ds \\
&\le&\frac12 \int_{s=0}^t e^{\Gamma_s}\ddot\psi_\eta(\nu
^{N,N'}_s)\bigl\{
|\nu^{N,N'}_s| + \bigl|Y^N_s-Y^N_{\vrho_N(s)}\bigr|
+ \bigl|Y^{N'}_s-Y^{N'}_{\vrho_{N'}(s)}\bigr|\bigr\} \,ds
\\
&\le&\frac1{2\ln\eta^{-1}} \int_{s=0}^t e^{\Gamma_s}\biggl\{1 + \frac
{1}{\eta}\bigl|Y^N_s-Y^N_{\vrho_N(s)}\bigr| + \frac{1}{\eta
}
\bigl|Y^{N'}_s-Y^{N'}_{\vrho_{N'}(s)}\bigr|\biggr\} \,ds \\
&\le&\frac1{4\ln\eta^{-1}} \int_{s=0}^t \biggl\{ e^{2\Gamma_s} + \biggl\{1 +
\frac{1}{\eta}\bigl|Y^N_s-Y^N_{\vrho_N(s)}\bigr|
+ \frac{1}{\eta}
\bigl|Y^{N'}_s-Y^{N'}_{\vrho
_{N'}(s)}\bigr|\biggr\}^2 \biggr\} \,ds \\
&\le&\frac1{4\ln\eta^{-1}} \int_{s=0}^t \biggl\{ e^{2\Gamma_s} + 3 +
\frac
{3}{\eta^2}\bigl|Y^N_s-Y^N_{\vrho_N(s)}\bigr|^2 
+ \frac{3}{\eta^2}\bigl|Y^{N'}_s-Y^{N'}_{\vrho
_{N'}(s)}\bigr|^2\biggr\} \,ds, \\
A^{2,N,N'}_t &=& \frac12 \int_{s=0}^t \ddot\psi_\eta(\nu
^{N,N'}_s)\bigl\{
\bigl(\bigl(Y^N_{\vrho^N(s)} +Z_s\bigr)\vee0\bigr) 
-\bigl(\bigl(Y^{N'}_{\vrho^{N'}(s)} +Z_s
\bigr)\vee
0\bigr)\bigr\}^2 \,ds\\
&\le&\frac12 \int_{s=0}^t \ddot\psi_\eta(\nu^{N,N'}_s)
\bigl|Y^N_{\vrho
_N(s)}-Y^{N'}_{\vrho_{N'}(s)}\bigr|^2 \,ds \\
&\le&\frac32 \int_{s=0}^t \ddot\psi_\eta(\nu^{N,N'}_s)\bigl\{
|\nu
^{N,N'}_s|^2 + \bigl|Y^N_s-Y^N_{\vrho_N(s)}\bigr|^2
+ \bigl|Y^{N'}_s-Y^{N'}_{\vrho_{N'}(s)}\bigr|^2\bigr\} \,ds
\\
&\le&\frac{3}{2\ln\eta^{-1}} \int_{s=0}^t \biggl\{\eta^{1/2} + \frac
{1}{\eta}\bigl|Y^N_s-Y^N_{\vrho_N(s)}\bigr|^2 + \frac{1}{\eta
}
\bigl|Y^{N'}_s-Y^{N'}_{\vrho_{N'}(s)}\bigr|^2\biggr\} \,ds.
\end{eqnarray*}
In the bound on $A^{1,N,N'}$, we have used Young's inequality, and in
the bound on $A^{2,N,N'}$ we have used the fact that the support of
$\ddot\psi_\eta$ is contained in $[0,\sqrt{\eta}]$.
Collecting things together, we have that there is a $\KK>0$ such that
\begin{eqnarray*} \BE[A^{1,N,N'}_t]&\le&\frac{\KK}{\ln\eta^{-1}}\biggl\{
1 +
\frac{1}{N\eta^2} + \frac{1}{N' \eta^2}\biggr\},\\
\BE[A^{2,N,N'}_t]&\le&\frac{\KK}{\ln\eta^{-1}}\biggl\{\eta^{1/2}+ \frac
{1}{N\eta} + \frac{1}{N'\eta}\biggr\}
\end{eqnarray*}
for all $t\in[0,T]$.
Thus,
\[
\lim_{N,N'\to\infty} \BE[|\nu^{N,N'}_t|] \le\sqrt{\eta} + \frac
{\KK
\sigma^2}{\ln\eta^{-1}} + \frac{\KK\beta^2\eta^{1/2}}{\ln\eta^{-1}}
\]
for all $t\in[0,T]$. Letting $\eta\searrow0$, we indeed get that
$\lim
_{N,N'\to\infty} \BE[|\nu^{N,N'}_t|]=0$.

We thus have that
\[
\mathop{\overline{\lim}}_{N,N'\to\infty} \BE
[|Y^N_t-Y^{N'}_t|] = 0.
\]
For any $p>1$, we also have by interpolation and (\ref{Eaa}) and (\ref
{Ebb}) that
\begin{eqnarray*}
\mathop{\overline{\lim}}_{N,N'\to\infty} \BE
[|Y^N_t-Y^{N'}_t|^p] & \le&
\mathop{\overline{\lim}}_{N,N'\to\infty} \sqrt{\BE
[|Y^N_t-Y^{N'}_t|]\BE
[|Y^N_t-Y^{N'}_t|^{2p-1}]} \\
& =&0.
\end{eqnarray*}
Thus, there is a solution $Y$ of the integral equation
\[
Y_t = \sigma\int_{s=0}^t e^{\Gamma_s/2}\sqrt{(Y_s+Z_s
)\vee
0}\,dW^*_s + \beta\int_{s=0}^t\bigl((Y_s +Z_s)\vee
0
\bigr)\,dV_s
\]
such that $\sup_{t\in[0,T]}\BE[|Y_t|^p]<\infty$ for all $T>0$ and
$p\ge1$. Setting $\bar Y_t \Def Z_t + Y_t$, we have that $\bar Y_t\in
\bigcap_{p\ge1}L^p$ and that
\[
\bar Y_t = Z_t+\sigma\int_{s=0}^t e^{\Gamma_s/2}\sqrt{\bar Y_s\vee
0}\,dW^*_s + \beta\int_{s=0}^t(\bar Y_s\vee0)\,dV_s.
\]

We claim that $\bar Y$ is nonnegative. For each $\eta>0$ we have that
\begin{eqnarray*} \psi_\eta(\bar Y_t)\chi_{\R-}(\bar Y_t)&=& \psi
_\eta
(\lambda_\circ)\chi_{\R-}(\lambda_\circ) + \frac{\sigma^2}{2}
\int
_{s=0}^t \ddot\psi_\eta(\bar Y_s)\chi_{\R-}(\bar Y_s)e^{\Gamma
_s/2}(\bar Y_s\vee0)\,ds \\
&&{}+ \frac{\beta^2}{2} \int_{s=0}^t \ddot\psi_\eta(\bar Y_s)\chi
_{\R
-}(\bar Y_s)(\bar Y_s\vee0)^2\,ds +\mart_t,
\end{eqnarray*}
where $\mart$ is a martingale.
Taking expectations and then letting $\eta\searrow0$, we have that
$\BE
[\bar Y_t^-]=0$.
We finally set $\lambda_t \Def e^{-\Gamma_t}\bar Y_t$. The claim follows.
\end{pf*}
\begin{pf*}{Proof of Lemma~\ref{Luniqueness}} Let $\lambda$ and
$\lambda'$ be two solutions of (\ref{ElambdaSDE}). Define $Y_t \Def
\lambda_te^{\Gamma_t}-Z_t$ and $Y'_t \Def\lambda'_t e^{\Gamma_t}-Z_t$.
Since $\lambda$
and $\lambda'$ are assumed to be nonnegative, $Y$ and $Y'$ satisfy
\begin{eqnarray*} Y_t &=& \sigma\int_{s=0}^t e^{\Gamma_s/2}\sqrt
{Y_s+Z_s}\,dW^*_s + \beta^S \int_{s=0}^t(Y_s+Z_s)\,dV_s,\\
Y'_t &=& \sigma\int_{s=0}^t e^{\Gamma_s/2}\sqrt{Y'_s+Z_s}\,dW^*_s +
\beta
^S \int_{s=0}^t(Y'_s+Z_s)\,dV_s.
\end{eqnarray*}
Set $\nu_t \Def Y_t-Y'_t$. For each $\eta>0$,
\[
|\nu_t| \le\psi_\eta(\nu_t) + \sqrt{\eta} = \sigma^2 A^1_t +
(\beta^S)^2A^2_t + \mart_t +\sqrt{\eta},
\]
where $\mart$ is a martingale and where
\begin{eqnarray*} A^1_t &=& \frac12 \int_{s=0}^t \ddot\psi_\eta(\nu
_s)e^{\Gamma_s}\bigl\{\sqrt{Y_s+Z_s}-\sqrt{Y'_s+Z_s}\bigr\}^2 \,ds \le\frac
1{\ln\eta^{-1}} \int_{s=0}^t e^{\Gamma_s} \,ds,\\
A^2_t &=& \frac12 \int_{s=0}^t \ddot\psi_\eta(\nu_s)\nu^2_s \,ds
\le\frac
{\eta^{1/2}}{\ln\eta^{-1}} t.
\end{eqnarray*}
Collecting things together, we have that
\[
\BE[|\nu_t|] \le\sqrt{\eta} + \frac{1}{\ln\eta^{-1}}\biggl\{\sqrt
{\eta}t
+ \int_{s=0}^t \BE[e^{\Gamma_s}] \,ds\biggr\}.
\]
Let $\eta\searrow0$ to get that $Y=Y'$. The claim follows.\vadjust{\goodbreak}
\end{pf*}

Let's next prove the needed macroscopic bound on the $\lambda^\NN$'s.
\begin{pf*}{Proof of Lemma~\ref{Lmacrobound}} For each $N\in\N$ and
$n\in\{1,2,\ldots, N\}$, define
\begin{eqnarray*} \Gamma^\NN_t &\Def&\alpha_\NN t + \beta^S_\NN
\gamma\int
_{s=0}^t X_s\,ds, \\
Z^\NN_t&\Def&\lambda_{N,n,\circ} +\alpha_\NN\bar\lambda_\NN\int
_{s=0}^t e^{\Gamma^\NN_s}\,ds+ \beta_\NN^C\int_{s=0}^t e^{\Gamma_s}\,dL^N_s
\end{eqnarray*}
and let $Y^\NN$ satisfy the equation
\begin{eqnarray*}
Y^\NN_t & = & \sigma_\NN\int_{s=0}^t e^{\Gamma^\NN_s/2}\sqrt{Y^\NN
_s+Z^\NN
_s}\,dW^n_s \\
&&{} + \eps_N \beta^S_\NN\int_{s=0}^t(Y^\NN_s +Z^\NN
_s)\,dV_s;
\end{eqnarray*}
then $\lambda^\NN_t = e^{\Gamma^\NN_t}(Y^\NN_t+Z^\NN
_t)$. We
calculate that
\begin{eqnarray*}
|\lambda^\NN_t|^p & \le &\tfrac12\{ e^{-2p\Gamma^\NN_t} + |Y^\NN
_t+Z^\NN
_t|^{2p}\}\\
& \le &\tfrac12\{ e^{-2p\Gamma^\NN_t} + 2^{2p-1}(|Y^\NN_t|^{2p} +
|Z^\NN_t|^{2p})\}.
\end{eqnarray*}
From Lemma~\ref{LTechnicalBounds1}, we have that
\[
\mathop{\sup_{0\le t\le T }}_{ N\in\N}\frac{1}{N}\sum_{n=1}^N\BE
[e^{2p\Gamma^\NN_t}] \quad\mbox{and}\quad \mathop{\sup_{
0\le t\le T }}_{ N\in\N}\frac{1}{N}\sum_{n=1}^N\BE[|Z^\NN
_t|^{2p}]
\]
are both finite.

For each $N\in\N$ and $n\in\{1,2,\ldots, N\}$, we compute that
\begin{eqnarray*} \BE[|Y^\NN_t|^{2p}] &=& p(2p-1)\biggl\{\sigma
_\NN
^2 \int_{s=0}^t \BE[|Y^\NN_s|^{2p-2}e^{\Gamma^\NN_s}
|Y^\NN
_s+Z^\NN_s|]\,ds \\
&&\hspace*{51.2pt}{}+ \eps_N^2 (\beta^S_\NN)^2 \int_{s=0}^t \BE
[|Y^\NN_s|^{2p-2}|Y^\NN_s+Z^\NN_s|^2]\,ds\biggr\}.
\end{eqnarray*}
To bound the integrals, we have that
\begin{eqnarray*}
&&
|Y^\NN_s|^{2p-2}e^{\Gamma^\NN_s}|Y^\NN
_s+Z^\NN
_s|\\
&&\qquad\le
\frac{1}{2p}e^{2p\Gamma^\NN_t} + \frac{p-1}{p}|Y^\NN_t|^{2p}
+ \frac{1}{2p}|Y^\NN_s+Z^\NN_s|^{2p} \\
&&\qquad
\le\frac{1}{2p}e^{2p\Gamma^\NN_t} + \frac{p-1}{2p}|Y^\NN_t|^{2p}
+ \frac{2^{2p-1}}{2p}\{|Y^\NN_s|^{2p}+
|Z^\NN
_s|^{2p}\},\\
&&|Y^\NN_s|^{2p-2}|Y^\NN_s+Z^\NN_s|^2 \\
&&\qquad\le\frac
{p-1}{p}|Y^\NN
_s|^{2p} + \frac{1}{p}|Y^\NN_s+Z^\NN_s|^{2p} \\[-2pt]
&&\qquad\le\frac{p-1}{p}|Y^\NN_s|^{2p}
+ \frac{2^{2p-1}}{p}\{|Y^\NN_s|^{2p}+|Z^\NN_s|^{2p}\}.
\end{eqnarray*}
Combining things together, we have that there is a $\KK>0$ such that
\begin{eqnarray*} \BE[|Y^\NN_t|^{2p}] &\le&\KK\{\sigma
_\NN
^2 + \eps_N^2 (\beta^S_\NN)^2\}\\[-2pt]
&&\hspace*{0pt}{}\times\biggl\{\int_{s=0}^t \BE
[|Y^\NN_s|^{2p}]\,ds + \int_{s=0}^t \BE[e^{2p\Gamma^\NN_s}]\,ds\\[-2pt]
&&\hspace*{101pt}{} +
\int
_{s=0}^t\BE[|Z^\NN_s|^{2p}]\,ds\biggr\}
\end{eqnarray*}
for all $N\in\N$ and $n\in\{1,2,\ldots, N\}$.
Using Assumption~\ref{ABounded} and averaging over $n$, we get the
claimed result.
\end{pf*}

\subsection{\texorpdfstring{Proof of Lemma \protect\ref{LbQDef}}{Proof of Lemma 4.1}}\label{SSBounds2}

Define a homeomorphism $\Phi$ of $C[0,\infty)$ as
\begin{eqnarray*} \Phi(q)(t) &\Def&
\mathop{\int_{\hat\mathsf
{p}=(\mathsf{p},\lambda
)\in\hat\PP}}_{
\mathsf{p}=(\alpha,\bar\lambda,\sigma,\beta^C,\beta^S)} \beta^C
\biggl[ \dot
b^\mathsf{p}(t)\lambda+ \int_{r=0}^t\dot b^\mathsf{p}(t-r)\{ q(r)+
\alpha\bar
\lambda\} \,dr\biggr] \\[-2pt]
&&\hspace*{61pt}{} \times\exp\biggl[-b^\mathsf{p}(t)\lambda- \int_{r=0}^t
b^\mathsf{p}(t-r)
\{ q(r) + \alpha\bar\lambda\} \,dr
\biggr]\\[-2pt]
&&\hspace*{61pt}{}\times\pi(d\mathsf{p})\Lambda_\circ(d\lambda)
\end{eqnarray*}
for all $q\in
C[0,\infty)$ and $t\ge0$. Note that since $b$, $q$ and $\lambda$ are
all nonnegative,
\[
0\le\exp\biggl[-b^\mathsf{p}(t)\lambda- \int_{r=0}^t b^\mathsf
{p}(t-r)\{ q(r) +
\alpha\bar\lambda\} \,dr\biggr]\le1.
\]
We can then set up a recursion; we want
to solve $Q = \Phi(Q)$. Note that there is a $\KK>0$ such that
\[
|\Phi(q)(t)|\le\KK\int_{s=0}^t q(r)\,dr
\]
for all nonnegative $q\in C[0,\infty)$.

For any $q_1$ and $q_2$ in $C[0,\infty)$, we have that
\[
\Phi(q_1)(t) -\Phi(q_2)(t) = \Gamma^a_t(q_1,q_2) + \Gamma^b_t(q_1,q_2),
\]
where
\begin{eqnarray*}
\hspace*{-4pt}&&\Gamma^a_t(q_1,q_2) \\[-3pt]
\hspace*{-4pt}&&\qquad\Def \int_{s=0}^t \biggl\{\int_{\theta=0}^1 \mathop
{\int
_{\hat\mathsf{p}=(\mathsf{p},\lambda)\in\hat\PP}}_{
\mathsf{p}=(\alpha,\bar\lambda,\sigma,\beta^C,\beta^S)} \beta^C
\dot b^\mathsf{p}
(t-s)\{ q_1(s)-q_2(s)\} \\[-3pt]
\hspace*{-4pt}&&\hspace*{25.4pt}\qquad\quad{} \times\exp\biggl[-b^\mathsf{p}(t)\lambda\\[-3pt]
\hspace*{-4pt}&&\hspace*{91.6pt}{} - \int_{r=0}^t
b^\mathsf{p}
(t-r)\bigl[ \bigl\{ q_2(r) + \theta\bigl(q_1(r) -q_2(r)\bigr)\bigr\}
+ \alpha\bar
\lambda\bigr] \,dr \biggr]\\[-3pt]
\hspace*{-4pt}&&\qquad\quad\hspace*{215.7pt}{}\times\pi(d\mathsf{p})
\Lambda_\circ(d\lambda
)\,d \theta\biggr\}
\,ds,\\[-3pt]
\hspace*{-4pt}&&
\Gamma^b_t(q_1,q_2) \\[-3pt]
\hspace*{-4pt}&&\qquad\Def -\int_{s=0}^t \biggl\{\int_{\theta=0}^1 \mathop
{\int
_{\hat\mathsf{p}=(\mathsf{p},\lambda)\in\hat\PP}}_{
\mathsf{p}=(\alpha,\bar\lambda,\sigma,\beta^C,\beta^S)} \beta^C
\biggl\{\dot
b^\mathsf{p}(t)\lambda
+ \int_{r=0}^t\dot b^\mathsf
{p}(t-r)\\[-3pt]
\hspace*{-4pt}&&\qquad\quad\hspace*{142pt}{}\times\bigl[ \bigl\{
q_2(r) + \theta\bigl(q_1(r)-q_2(r)\bigr)\bigr\}+ \alpha\bar\lambda\bigr] \,dr\biggr\}
\\[-3pt]
\hspace*{-4pt}&&\qquad\quad\hspace*{28pt}{} \times\bigl\{ b^\mathsf{p}(t-s)\bigl(q_1(s)-q_2(s)\bigr)\bigr\} \\[-3pt]
\hspace*{-4pt}&&\qquad\quad\hspace*{28pt}{} \times\exp\biggl[ -b^\mathsf{p}(t)\lambda\\
\hspace*{-4pt}&&\qquad\quad\hspace*{59pt}{} - \int_{r=0}^t
b^\mathsf{p}
(t-r)\bigl[ \bigl\{ q_2(r) + \theta\bigl(q_1(r)-q_2(r)\bigr)\bigr\}
+ \alpha\bar\lambda\bigr] \,dr
\biggr] \\[-3pt]
\hspace*{-4pt}&&\qquad\quad\hspace*{26.7pt}\hspace*{189pt}{} \times\pi(d\mathsf{p})\Lambda_\circ(d\lambda)\,d\theta\biggr\} \,ds.
\end{eqnarray*}
Standard techniques from Picard iterations give us the result.

\section*{Acknowledgments}

R. B. Sowers would like to thank the Departments of Mathematics and
Statistics of Stanford University for their hospitality in the Spring
of 2010 during a sabbatical stay. The authors are grateful to Thomas
Kurtz for some insight into the literature on mean field models, and to
Michael Gordy and participants of the 3rd SIAM Conference on Financial
Mathematics and Engineering in San Francisco and the 2010 Annual
INFORMS Meeting in Austin for comments.



\printaddresses

\end{document}